\documentclass[10pt]{emulateapj}

\usepackage{comment}

\usepackage{graphicx}
\usepackage{color}
\usepackage{multirow}
\usepackage{url}

\newcommand{\LIGO}{LIGO}
\newcommand{\VIRGO}{Virgo}

\newcommand{\SG}{SG}
\newcommand{\G}{G}
\newcommand{\WNB}{WNB}
\newcommand{\BBH}{BBH}

\newcommand{\LIB}{LIB}
\newcommand{\cWB}{cWB}

\newcommand{\rev}[1]{#1}
\newcommand{\Rev}[1]{#1}

\newcommand{\rl}{\raggedleft}

\begin{document}

\title{
Localization of \rev{short duration} gravitational-wave transients with 
the early advanced LIGO and Virgo detectors
}

\author{
Reed Essick\altaffilmark{1}, Salvatore Vitale\altaffilmark{1}, 
Erik Katsavounidis\altaffilmark{1}, Gabriele Vedovato\altaffilmark{2}, 
Sergey Klimenko\altaffilmark{3}
}

\altaffiltext{1}{MIT LIGO Laboratory} 
\altaffiltext{2}{INFN Padova}
\altaffiltext{3}{University of Florida}

\begin{abstract}

The Laser Interferometer Gravitational wave Observatory 
(LIGO) and Virgo, advanced ground-based gravitational-wave 
detectors, will begin collecting science data in 
2015. With first detections expected to follow, it is 
important to quantify how well generic gravitational-wave 
transients can be localized on the sky. This is crucial for 
correctly identifying electromagnetic counterparts as well 
as understanding gravitational-wave physics and source 
populations. We present a study of sky localization 
capabilities for two search and parameter estimation algorithms:
\emph{coherent WaveBurst}, a \rev{constrained} likelihood algorithm operating in close 
to real-time, and \emph{LALInferenceBurst}, a Markov chain 
Monte Carlo parameter estimation algorithm developed to 
recover generic transient signals with latency of a few 
hours. Furthermore, we focus on the first few years of the 
advanced detector era, when we expect to only have two 
(2015) and later three (2016) operational detectors, all 
below design sensitivity. These detector configurations can 
produce significantly different sky localizations, which we 
quantify in detail. We observe a clear improvement in 
localization of the average detected signal when progressing 
from two-detector to three-detector networks, as expected. Although 
localization depends on the waveform morphology, approximately 
50\% of detected signals would be imaged after observing 
100-200 deg$^2$ in 2015 and 60-110 deg$^2$ in 2016, 
although knowledge of the waveform can reduce this to as
little as 22 deg$^2$. This is the first comprehensive study 
on sky localization capabilities for generic transients 
of the early network of advanced LIGO and Virgo detectors, 
including the early LIGO-only two-detector configuration.
\end{abstract}

\maketitle
\section{Introduction}

Advanced ground-based gravitational-wave detectors, such as 
the two advanced LIGO observatories 
[~\cite{Harry:2010}] and advanced Virgo
[~\cite{Accadia:2012}],
will begin collecting data as early as 2015. 
Although the detectors will not operate at design sensitivity 
initially, they will operate with enough sensitivity to 
possibly detect the first gravitational-wave transients 
[~\cite{Aasi:2013}]. This promises many scientific boons, 
and accurate waveform reconstruction and parameter estimation
will be key in extracting as much information as possible from these
detections.
In particular, accurate measurements 
of the sources' positions on the sky can help determine
their populations, their distributions, and possible formation 
mechanisms [~\cite{Dominik:2012,Kelley:2010,Krzysztof:2014}]. 
Furthermore, accurate sky localization will help electromagnetic 
follow-up to gravitational-wave transients, which may bring 
gravitational-wave observations into astrophysical and 
cosmological context. \Rev{This has been carefully studied for 
some binary systems [~\cite{Singer:2014}], and possible counterparts
have been proposed [~\cite{Metzger:2011,Barnes:2013}]}.

Searches for gravitational-wave transients are well motivated 
astrophysically. Among them, gravitational waves generated
by compact binary systems are the best understood, with well 
studied and modeled waveforms. Therefore, searches targeting 
compact binary systems employ matched filtering techniques 
[~\cite{PhysRevD.85.082002}]. 
\rev{
Although significant effort has been invested in
analytical and numerical studies of expected waveforms from
compact binaries[~\cite{Ajith:2005,Ajith:2011,Damour:2008,
Hannam:2010,Sturani:2010,Aasi:2014a,Cannon:2013}], some uncertainties
still exist, particularly in binary black hole systems with spin
or large eccentricity. 
Several of the anticipated transient sources come with only 
poorly understood or phenomenological gravitational waveforms, 
such as gravitational radiation from core-collapse supernovae 
[~\cite{Ott:2009}]. These waveforms are typically extremely 
difficult to simulate, and in the case of supernovae, may be 
subject to stochastic processes that make templated searches 
difficult. 
Other transient gravitational-wave
sources include pulsar glitches,
starquakes associated with magnetars, and cosmic string cusps
[~\cite{Abadie:2012a}].
}
In addition, there is always the possibility of completely 
unanticipated signals from currently unknown sources. Generic
transient searches that make only minimal assumptions on the 
signal's morphology (waveforms, polarizations) are well suited 
to detect such sources and, in this way, complement
matched-filtering approaches [~\cite{Abadie:2012a}].

In this study, we focus on short-duration (less than one second) 
gravitational-wave transients, also referred to as bursts, 
which are typically un-modeled or poorly modeled. 
Moreover, we focus on source localization only, rather than waveform
reconstruction,
and attempt to quantify the expected uncertainties produced by
analysis of gravitational wave data in the early advanced detector era.
In order to assess the performance of our algorithms, we use
families of ad hoc waveforms as a proxy for what may accompany 
astrophysical events [~\cite{Abadie:2012a}]. 

\rev{
We used three ad-hoc signal morphologies and along with Binary Black Hole (\BBH) coalescences to 
explore a wide range of possible signals detectable by generic burst searches. 
The ad-hoc morphologies scan the signal phase-space
\Rev{
with templates possessing both small and large 
time-frequency areas that
}
 span the entire sensitive frequency band of the instruments. 
Such waveforms should approximate the localization of signals from possible 
burst sources like core-collapse \Rev{s}upernovae. Depending on the mechanism [~\cite{Ott:2009}], their waveforms may 
resemble high-$Q$ sine-Gaussian signals (like in the acoustic mechanism [~\cite{Burrows:2006,Burrows:2007,Ott:2006}]), millisecond-scale 
Gaussian-like peaks (as in simulations of the rotating collapse and bounce in 
CCSNe [~\cite{Dimmelmeier:2008}]) or even white-noise bursts (if turbulent convection takes place [~\cite{Ott:2009}]).
Recent studies have focused on inferring the explosion mechanism from gravitational waveforms [~\cite{Logue:2012}], although they did not address localization.
}
\rev{Furthermore, using a range of morphologies } allows us to characterize 
localization for generic signals which may come from unanticipated sources.
This work is analogous to a recent study focusing on binary neutron star (BNS) 
coalescences [~\cite{Singer:2014}]. Signals from such systems typically 
have longer signal durations with well known \rev{broadband} waveforms and are targeted
more optimally with matched filter searches [~\cite{PhysRevD.85.082002}]. 

Accurately localizing gravitational wave signals can shed light 
on the sources' distribution across the sky and possibly
lead to identification of counterparts throughout the 
electromagnetic spectrum. \rev{Again, this has been carefully 
considered for a few scenarious} [~\cite{Feng:2014,Evans:2012,0067-0049-211-1-7}]
\rev{but is difficult to address for un-modeled bursts. However, accurate}
characterization of gravitational wave localization will naturally
inform any electromagnetic follow-up effort. 
[
There exist 
a host of possible counterparts to generic gravitational 
wave bursts. For instance, core-collapse \Rev{s}upernovae in the 
local universe are expected to produce detectable gravitational 
radiation [~\cite{Ott:2009}] and will have bright counterparts 
throughout the electromagnetic spectrum as well as in 
low-energy neutrinos [~\cite{Schol:2012}]. Superconducting 
cosmic string cusps are expected to produce both gravitational 
[~\cite{Aasi:2014b}] and electromagnetic radiation [~\cite{Vachaspati:2008}]. 
\BBH\ systems may or may not produce electromagnetic counterparts, 
depending on the system's environment.
For \BBH\ systems in a clean environment, gravitational-wave data
may be the only way to study these systems.
Furthermore, coincident electromagnetic observations for bursts from
unknown sources will be invaluable in determining the associated
physical system.
Regardless of the source of gravitational 
radiation, electromagnetic and neutrino observations \rev{may} place 
the event in an astrophysical context. 

Although several algorithms provide source localization estimates, 
we focus on Coherent WaveBurst (\cWB) [~\cite{Klimenko:2005,Klimenko:2008}], 
a \rev{constrained} likelihood algorithm \rev{(Section \ref{section:cWB})}, and LALInferenceBurst (\LIB) 
[~\cite{2013PhRvD..88f2001A,LALInference,LALInferenceCode}], 
a Markov chain Monte Carlo (MCMC) parameter estimation algorithm \rev{(Section \ref{section:LIB})}.
Previous sky localization studies for un-modeled bursts used an 
earlier version of \cWB\ 
and investigated networks with three or more detectors 
[~\cite{Klimenko:2011,Aasi:2013,Markowitz:2008,Abadie:2012c}]. 
Furthermore, these studies typically focused on a few sample 
waveforms with a few fixed parameter values. This includes 
characterizing algorithmic performance as a function of 
injection amplitude, for example. 
We focus on ensemble averages computed over a population of 
events with randomly selected parameters and with the expected 
detector configurations for the first two years of the advanced 
detector era. In particular, we generate an astrophysical 
population of generic burst events 
\Rev{
that extends beyond the detectors' sensitivity limits.
}
This characterizes the 
localization capabilities for typical events, and models 
the relative frequency of ``loud'' signals versus the more 
common ``quiet'' signals. This population yields estimates that 
describe a ``typical expected event''
from gravitaional-wave detectors.
We present an analysis of sky 
localization during the transition from two detectors 
(LIGO-Hanford and LIGO-Livingston in 2015) to three detectors 
(LIGO-Hanford, LIGO-Livingston and Virgo in 2016) with expected 
noise curves made available by the LIGO and Virgo collaborations 
[~\cite{Aasi:2013}].

Unlike many electromagnetic observations, gravitational-wave
source position uncertainties are very large, typically larger 
than 100 deg$^2$. Therefore, gravitational-wave searches 
produce probability distributions over the sky, rather than 
single locations, from which meaningful quantities are derived. 
These probability distributions can have very complicated
 shapes, including severe fragmentation and spatially separated 
support. A thorough understanding of these distributions can
inform the design of follow-up programs as well as the choice
of which events should be pursued.

This paper is organized as follows: Section 
\ref{section:Data preparation and injection generation} describes 
the simulated noise and gravitational waveforms we use in 
this study; Section \ref{section:sky localization pipelines} 
briefly describes the two algorithms we use; Section 
\ref{section:results} discusses the observed localization 
capabilities of the two pipelines over the same set of detected 
signals; Section \ref{section:systematics} describes some 
systematics associated with these algorithms and we conclude 
in Section \ref{section:conclusion}.

\section{Data preparation}
\label{section:Data preparation and injection generation}

\subsection{Noise}

\begin{figure}
  \includegraphics[width=0.5\textwidth]{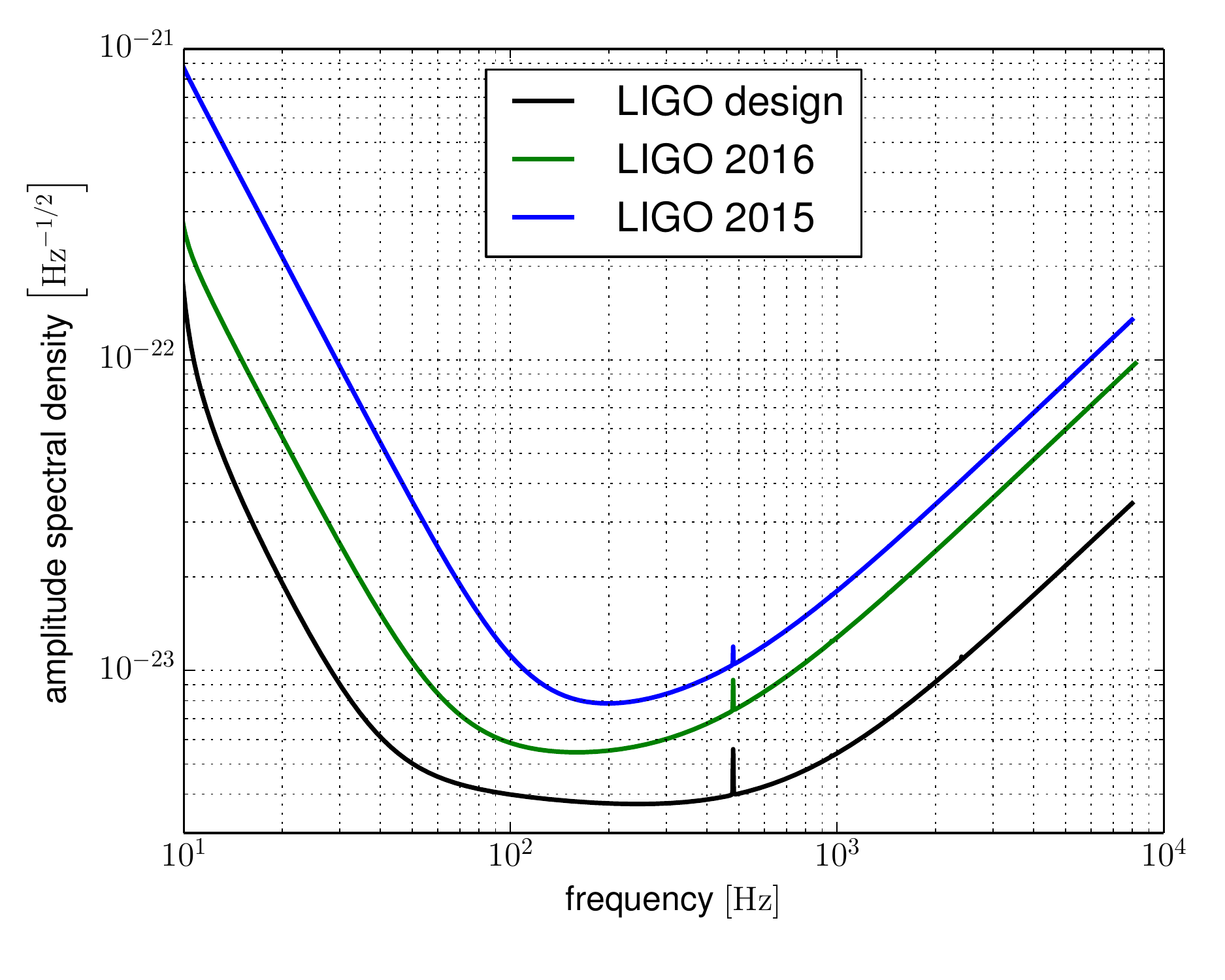}
  \includegraphics[width=0.5\textwidth]{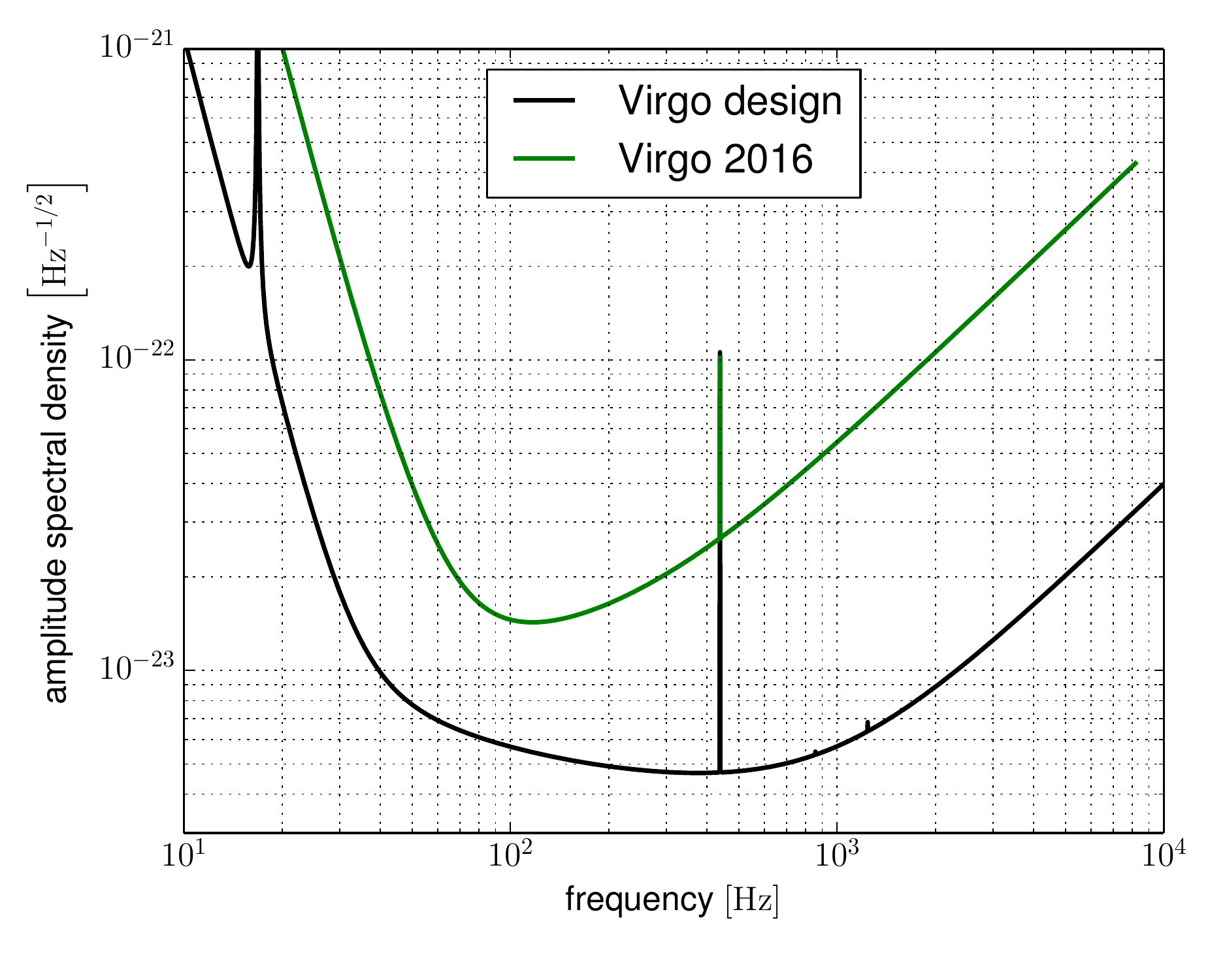}
  \caption{Expected amplitude strain noise for advanced detectors 
in 2015, 2016 and at design sensitivity. The resonances correspond 
to suspension ``violin'' modes near 500 Hz and 
``bounce'' modes near 15 Hz. The bounce mode is within the seismic 
wall for the LIGO curves.}
  \label{fig:noise curves}
\end{figure}

We use simulated stationary Gaussian noise throughout this study. 
Expected noise curves for the two LIGO detectors and Virgo 
are shown in Figure \ref{fig:noise curves}, which 
plots the curves for LIGO in 2015, 2016 and at design 
sensitivity as well as Virgo curves in 2016 and at design 
sensitivity. The 2015 and 2016 curves were chosen as the 
geometric mean of the optimistic and pessimistic estimates in 
[~\cite{Aasi:2013}], and the actual improvement in the noise 
curves will depend on the commissioning of the detectors. While 
these curves may not be realized exactly, they provide a good 
estimate for the Gaussian noise expected in the advanced 
detector era.

Another important source of noise, particularly for burst 
searches, is \emph{non-Gaussian} in nature. These non-Gaussian
 artifacts (``glitches'') form a long tail at high 
amplitudes, reducing the sensitivity of searches 
[~\cite{LVC-DQ:2014}].
Because we 
simulate only Gaussian noise, we do not have glitches present 
in our data. However, the presence or absence of glitches 
will not affect our study. For a given event, glitches 
affect \Rev{a gravitational-wave} observer's confidence that this particular event 
is of astrophysical origin, but will not affect the 
pipeline's ability to localize true astrophysical events. 

Furthermore, we do not simulate realistic detector livetimes
 because we do not attempt to make a statement about detection 
rates. Instead, we use one month of continuous simulated 
noise for each of the 2015 and 2016 runs. 

\subsection{Injections}

Several signal morphologies were used in an attempt to simulate 
a wide range of possible gravitational-wave transients. While 
these are not meant to be examples of actual astrophysical 
signals (with the exception of compact binary systems),
 historically several exemplar waveforms have been used to 
approximate generic burst events 
[~\cite{Aasi:2013,Klimenko:2011,Markowitz:2008}].
We focus on generic waveform morphologies because they allow 
us to test our ability to localize very different signals without
specializing to a specific source.
In this study we use four signal morphologies: Sine-Gaussian 
(\SG), Gaussian (\G), White-Noise Bursts (\WNB), and Binary 
Black Hole approximants (\BBH) with spins (anti-)aligned with 
the orbital angular momentum. 
The parameter ranges were chosen as reasonable given the 
typical frequencies probed by burst searches ($32-2048\,
\mathrm{Hz}$) as well as the expected noise curves. The 
parameters were drawn independently, and Table \ref{table:params} 
lists the exact values used.

In particular, we distribute our injections as if they
 were astrophysical, i.e., uniform in comoving volume. In 
addition, the quietest signals were chosen to be just below 
the detector's maximum sensitivity. This ensures that the 
signals recovered were limited by the detector's sensitivity 
rather than an artificial threshold. 
All populations were distributed uniformly over the sky and 
regularly spaced in time. 

As mentioned above, we distribute our injections uniformly 
in comoving volume. This is done using 
standard $\Lambda$CDM cosmology ($\Omega_m=0.3$, 
$\Omega_\Lambda=0.7$). If we take \Rev{s}uper\rev{n}ova as typical energy scales for 
un-modeled bursts, an optimistic upper limit for the energy emitted
as gravitational waves is $E_{GW}\sim 10^{-4}\,M_\odot c^2$ [~\cite{Ott:2006}]. 
If we assume a energy scale ten times larger is associated with 
and isotropically radiated \SG\ with $f_o=200$ Hz, this yields a horizon distance
of $\sim3.8$ Mpc with advanced LIGO design sensitivity. At this distance, the difference
between volume and comoving-volume is negligible ($\sim 0.1\%$). Therefore, 
we distribute \SG, \G, and \WNB\ signals uniformly in volume.

Furthermore, because we do not have an exact energy 
scale for generic (un-modeled) transient events, it is difficult 
to compute a distance. We expect the signal amplitude to scale 
inversely with the luminosity distance, and this can be used to 
define a distribution over $h_{rss}^2 = 
\int\mathrm{d}t\,\left( h_+^2 + h_\times^2 \right)$ given a distribution
over distance. A derivation is provided in the Appendix, but we can 
model a uniform-in-volume distribution as

\begin{equation}
p(D_L) \propto D_L^2 \Rightarrow p(h_{rss}) \propto h_{rss}^{-4}
\end{equation}

\BBH\ systems should be detectable at several Gpc, 
and the difference between volume and comoving volume is 
non-trivial here ($\ge 70\%$). For the \BBH\ signals, we 
have a well defined distance and distribute the signals 
uniformly in comoving volume. 

\subsubsection{Sine-Gaussian waveforms}

Sine-Gaussian (\SG) waveforms have historically been used by
 the \LIGO\ and \VIRGO\ Collaborations to simulate generic 
bursts [~\cite{Abadie:2012a}]. We define our \SG\ waveforms 
according to Equations \ref{equation:sg waveform hx} and 
\ref{equation:sg waveform h+} . 
$f_o$ is the central frequency of the Sine-Gaussian; $\tau$ 
is the width in the time domain.
$\alpha$ controls the relative weights between the two 
polarizations. This is equivalent to choosing the coordinate 
system in the wave-frame relative to the Earth-fixed detector 
frame.

\begin{widetext}
  \begin{eqnarray}
  h_{\times}(t) = & \sin\left(\alpha\right) \frac{h_{rss}}{\sqrt{Q(1-\cos\left(2\phi_o\right)e^{-Q^2})/4f_o\sqrt{\pi}}} \sin\left( 2\pi f_o (t-t_o) + \phi_o\right) e^{-(t-t_o)^2/\tau^2} \label{equation:sg waveform hx} \\
  h_{+}(t)      = & \cos\left(\alpha\right) \frac{h_{rss}}{\sqrt{Q(1+\cos\left(2\phi_o\right)e^{-Q^2})/4f_o\sqrt{\pi}}} \cos\left( 2\pi f_o (t-t_o) + \phi_o\right) e^{-(t-t_o)^2/\tau^2} \label{equation:sg waveform h+}
  \end{eqnarray}
\end{widetext}

\subsubsection{Gaussian waveforms}

We also inject Gaussian envelops in the time domain (\G), 
defined by Equations \ref{equation:g waveform hx} and 
\ref{equation:g waveform h+}. These can be considered as limiting 
cases of \SG\ waveforms in which $f_o\rightarrow0$. However, 
removing the oscillatory component means the frequency domain
 waveform is a Gaussian centered about $f=0$, and the signal
 is detected essentially by the Gaussian's wings. Because 
the seismic wall in the noise spectra at low frequencies is 
very steep, small changes in Gaussian width can significantly 
affect detectability. This and the lower bound on signal 
duration from the pipeline's sampling rate determined the 
injection population's parameter ranges.

\begin{eqnarray}
h_{\times}(t) = & \sin\left(\alpha\right) \frac{h_{rss}}{\sqrt{\tau}}\left(\frac{2}{\pi}\right)^{1/4} e^{-(t-t_o)^2/\tau^2} \label{equation:g waveform hx} \\
h_{+}(t)     = & \cos\left(\alpha\right) \frac{h_{rss}}{\sqrt{\tau}}\left(\frac{2}{\pi}\right)^{1/4} e^{-(t-t_o)^2/\tau^2} \label{equation:g waveform h+}
\end{eqnarray}

\subsubsection{White-Noise Burst waveforms}

Perhaps one of the most generic waveforms we investigate is 
the white-noise burst (\WNB), defined by Equations 
\ref{equation:wnb waveform hx} and \ref{equation:wnb waveform h+}. 

\begin{widetext}
  \begin{eqnarray}
  h_{\times}(t) \propto & e^{-(t-t_o)^2/\tau^2} \int e^{-2\pi i f t} [\Theta(f-f_{min}) - \Theta(f-f_{max})]\, \mathrm{d}f \label{equation:wnb waveform hx} \\
  h_{+}(t)      \propto & e^{-(t-t_o)^2/\tau^2} \int e^{-2\pi i f t} [\Theta(f-f_{min}) - \Theta(f-f_{max})]\, \mathrm{d}f \label{equation:wnb waveform h+}
  \end{eqnarray}
\end{widetext}

The flat frequency domain component is randomly drawn 
from Gaussian white noise rather than a truly flat curve. All 
parameters besides $h_{rss}$ are drawn first, including this 
randomly sampled frequency domain waveform, and then the 
amplitude is scaled appropriately to obtain the desired 
$h_{rss}$. These signals are meant to simulate an excess 
of power randomly distributed within some frequency band 
and localized in time.

\subsubsection{Binary Black Hole waveforms}

We simulate the inspiral of massive binary systems because 
they coalesce at relatively low frequencies. This means that 
the signal is relatively compact in the frequency domain, 
and generic burst searches can more easily detect these signals 
compared to lighter systems. We expect burst 
searches \rev{to} have sensitivity to Binary Black Hole (\BBH) 
coalescence at cosmological distances.

We use Inspiral-Merger-Ringdown (IMR) phenomenological 
approximants to model \BBH\ coalescence 
[~\cite{Ajith:2011,Hannam:2010}]. These waveforms are constructed 
by stitching analytic post-Newtonian expansions, accurate to 
3.5 PN orders, with numerical-relativity results for merger 
and analytic quasi-normal modes for ringdown. Typically, the inspiral portion of the waveform 
is known much more accurately than the merger and ringdown. 
However, our \BBH\ signals contain massive components and 
their mergers occur within the detector's sensitive band. 

Importantly, our simulated waveforms also incorporate the 
affects of spin-orbit coupling. We focus on spins (anti-)aligned 
with the orbital angular momentum, so there is no 
spin-precession in these waveforms. There is still uncertainty 
about the efficiency of common envelop evolution and the 
relative importance of supernova kicks [~\cite{Krzysztof:2014}], 
and astrophysical \BBH\ systems may or may not have 
their spins (anti-)aligned with the orbital angular momentum. 
However, this is a reasonable assumption when characterizing 
our algorithms' performance.

We use a range of component masses consistent with stellar 
mass black holes. We also simulate a wide range of spin 
magnitudes for each object. 

\section{Localization pipelines}
\label{section:sky localization pipelines}

The goal of  sky localization of gravitational-wave transients 
is to construct a posterior probability distribution over
 the sky. 
We use two pipelines to localize signals: Coherent WaveBurst
 (\cWB) and LALInferenceBurst (\LIB). Each pipeline attempts
 to reconstruct the signal's sky position in a different way, 
which we briefly describe. 

\subsection{Coherent WaveBurst}
\label{section:cWB}

\rev{
Coherent WaveBurst (\cWB) is a data analysis algorithm for the detection of transient gravitational-wave signals (bursts) [~\cite{Klimenko:2005,Klimenko:2008}]. 
In \cWB, burst events are identified as excess power patterns, exceeding some threshold, in the time-frequency domain obtained via a wavelet transformation. 
Assuming Gaussian noise, \cWB\ combines data from multiple detectors to compute a constrained likelihood functional dependent on the source’s sky position.  
For networks with two or three detectors, strong degeneracies exist in the likelihood and \cWB\
applies several ad-hoc constraints to limit the signal space.
The constrained likelihood is maximized  over all possible gravitational-wave signals for each point in the sky, and several statistics are computed at each point. 
\cWB\ generates maps by combining these statistics, which are used to approximate posterior probability distribution over the sky [~\cite{Klimenko:2011}]. 
}
Previous studies [\cite{Aasi:2013, Klimenko:2011}] used an earlier version of \cWB. 
We present results from an updated version of the algorithm, referred to as the second-generation: \cWB-2G [~\cite{cWB2G-prep}]. 

Importantly, this study implements an \emph{effective prior} on the source position in Earth-fixed coordinates.
This is the first implementation of such prior for burst detection algorithms, which modulate the posterior with the detectors’ antenna patterns to incorporate the fact that quieter signals are more frequent than loud signals. 
It is, therefore, more likely a priori to detect signals from parts of the sky with large antenna patterns. 
A derivation is provided in the Appendix.

\rev{
\cWB\ uses two network constraints (regulators), which incorporate prior knowledge on how the network responds to generic gravitational-wave signals. 
The network response to a signal is constrained by the antenna patterns; this is used in \cWB's analysis to ``regulate'' reconstructed signal waveforms and reduce the algorithm’s sensitivity to non-Gaussian noise artifacts. 
The regulators modify the form of the likelihood functional
and can be thought of as non-trivial priors. 
}
In \cWB-2G, the regulators are controlled by the parameters $\delta$ and $\gamma$. 
$\delta$ controls the permissible ratios between the contributions to the likelihood from separate polarizations, and \rev{$\gamma$ acts as the lower bound on the correlation between the detectors [~\cite{cWB2G-prep}].}
While the regulators are not needed to reject background in our simulated Gaussian noise, they will be used in an actual observing run. 
Section \ref{section:algorithmic params} describes the exact parameters used.

With two detectors, our choice of regulator settings force \cWB\ to reconstruct
only a single polarization. Because the Hanford and Livingston
detectors are nearly aligned, they are effectively sensitive
to only a single polarization [~\cite{Klimenko:2011,Sutton:2010}]
and this is a reasonable approximation. The three-detector
regulators are almost, but not quite, turned off. We
discuss the features introduced by the regulators in Section \ref{section:systematics}.

While false-alarm rates will depend on the Gaussian and non-Gaussian
noise in an observing run, we chose detection thresholds for
\cWB\ that correspond to a false alarm rate of 1 \rev{per year} 
in historical non-Gaussian noise. This threshold may not be
high enough to claim a confident detection, but it is likely
that events satisfying such criterion will be of significant interest.

\cWB\ is a low-latency pipeline, typically run as on online
search. Posteriors are produced as part of the detection
pipeline, and are available within minutes
of recording the data.

\subsection{LALInferenceBurst}
\label{section:LIB}

LALInferenceBurst (\LIB) is a Bayesian Markov chain Monte 
Carlo parameter estimation algorithm designed to recover burst 
signals and estimate some key signal parameters, including sky 
position. \LIB\ is based on nested sampling and shares most 
of its libraries with LALInference, its counterpart for 
parameter estimation of compact binary coalesces (CBC) 
[~\cite{LALInference}]. A detailed description of nested 
sampling and its application to gravitational-wave parameter 
estimation can be found elsewhere 
[~\cite{2004AIPC..735..395S,2010PhRvD..81f2003V}].

The main difference between the CBC and Burst version of LALInference 
is that, while the CBC version filters the data using 
long waveforms 
that describe the signal emitted by 
compact binaries, \LIB\ uses a single \SG\ waveform.
\footnote{It is possible to also use Gaussians or other short 
waveforms, but we do not consider them in this study.}
This implies that \LIB\ can not perfectly match 
some of the simulated signals considered in this study, such as 
\WNB\ and \BBH. We expect \LIB\ to perform sub-optimally for 
these signals, while it should still be able to recognize that 
a coherent signal is present, and produce
useful sky localization information. We see that this is 
indeed the case.

\LIB\ has a larger computational cost than \cWB, and it 
typically cannot be run as a blind search algorithm. Instead, 
\LIB\ is run as a follow-up to pre-selected times, 
with typical latencies between hours and days. It provides 
flexibility in tuning computational cost and sensitivity. 
With the configuration used in this analysis, 50\% of the 
events were processed within two hours.
However, a detection decision can be reached with latencies 
of a few minutes.
For this study, \LIB\ was run on a subset of triggers detected 
by \cWB\ and was not used as a search pipeline [~\cite{LIB-prep}].
Beside the approximate time of an event, \LIB\
does not use any data products produced by \cWB. 

Because \LIB\ is a template based algorithm, we can put 
priors on the (relatively few) parameters that describe the 
template. In this work, we put a uniform-in-volume prior 
on the template amplitude, $p(h_{rss})\propto h_{rss}^{-4}$. 
This is a more direct way of incorporating the prior knowledge 
than the effective prior used with \cWB, but the information 
content is similar. The prior furthermore assumes that 
sources are uniformly distributed on the sky, and modulation 
with the antenna patterns is achieved only through the prior 
on signal amplitude. The prior on other signal parameters was 
flat with ranges larger than the injected ranges.
Section \ref{section:algorithmic params} describes the exact 
parameters used.

\section{Results}\label{section:results}

While localizing sources using gravitational-wave data alone
 is important, past studies have emphasized directing 
electromagnetic follow-up [~\cite{Abadie:2012c,Abadie:2012b}]. 
Therefore, the metrics used to evaluate localization at least 
tacitly assume some electromagnetic follow-up program. We 
present results for a few standard measures for our
 simulations of the early advanced detector era.

\begin{figure}
	\begin{center}
		\includegraphics[width=0.5\textwidth, clip=True, trim=0.25cm 1.00cm 0.25cm 2.60cm]{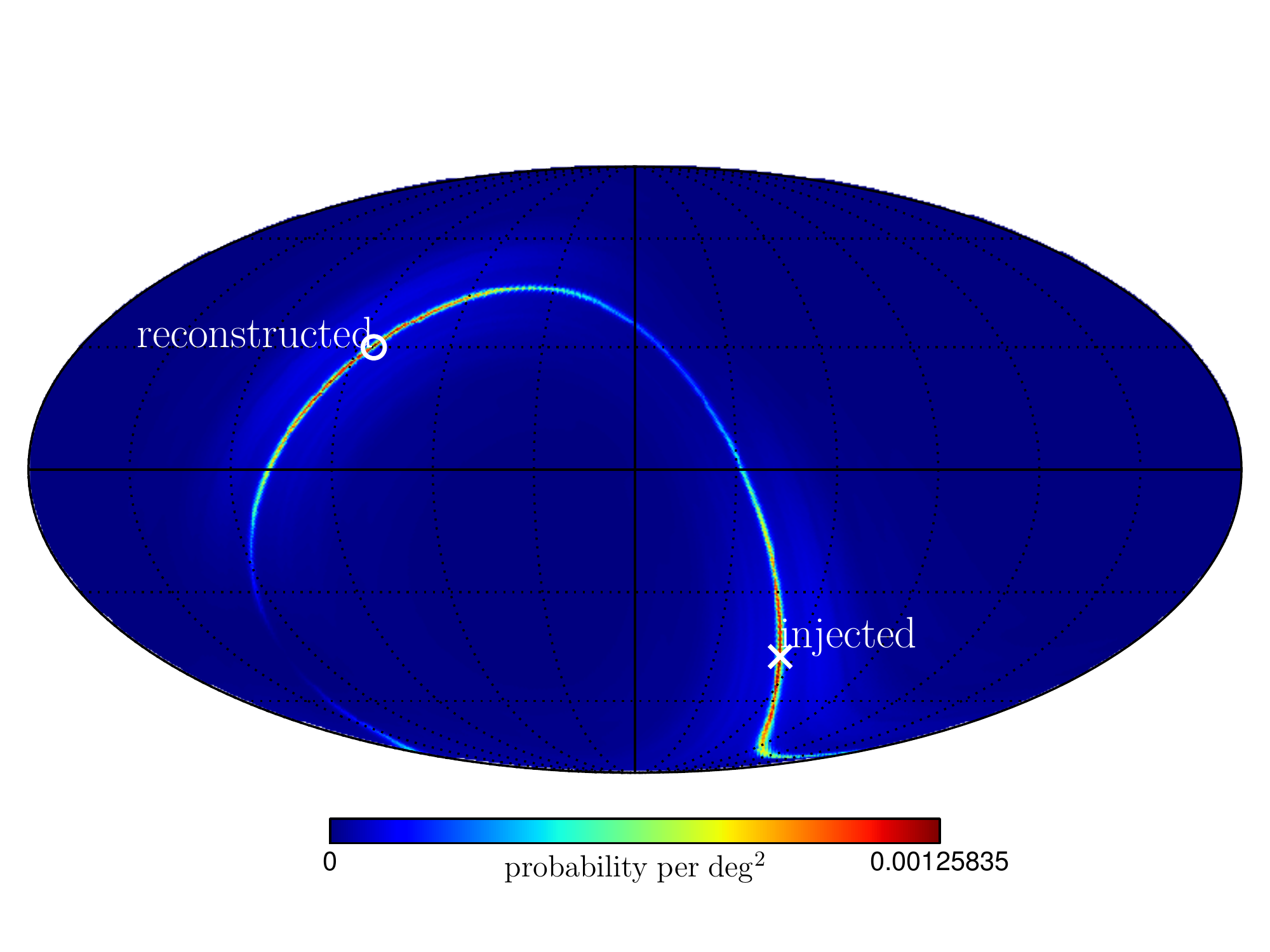} \\
		(a) sky map \\
	\end{center}
	\begin{minipage}{0.5\textwidth}
		\begin{center}
                	\includegraphics[width=\textwidth, clip=True, trim=0.25cm 1.00cm 0.25cm 2.60cm]{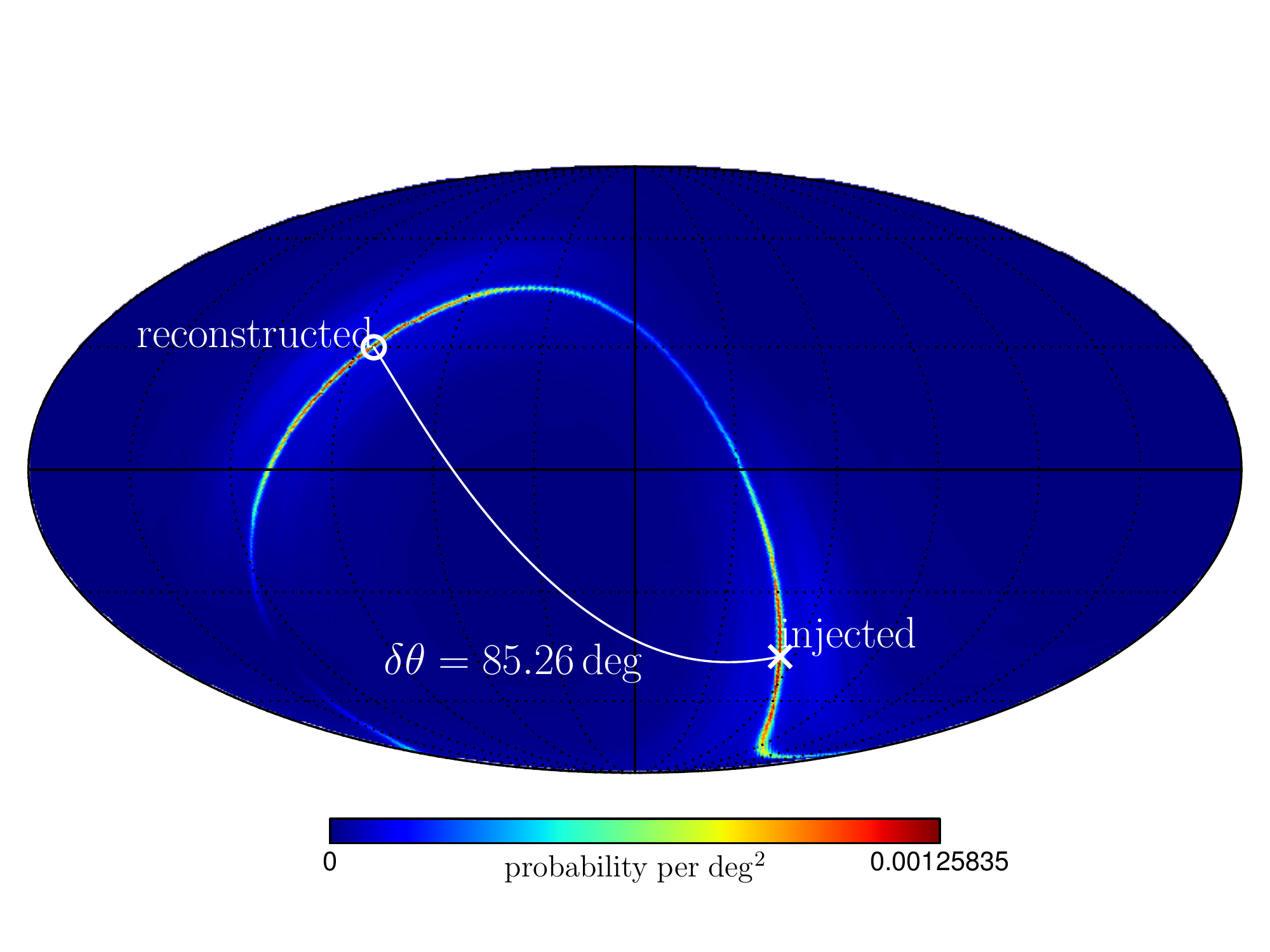} \\
			(b) $\delta \theta$ \\
		\end{center}
	\end{minipage}
	\begin{minipage}{0.5\textwidth}
		\begin{center}
 	               \includegraphics[width=\textwidth, clip=True, trim=0.25cm 1.00cm 0.25cm 2.60cm]{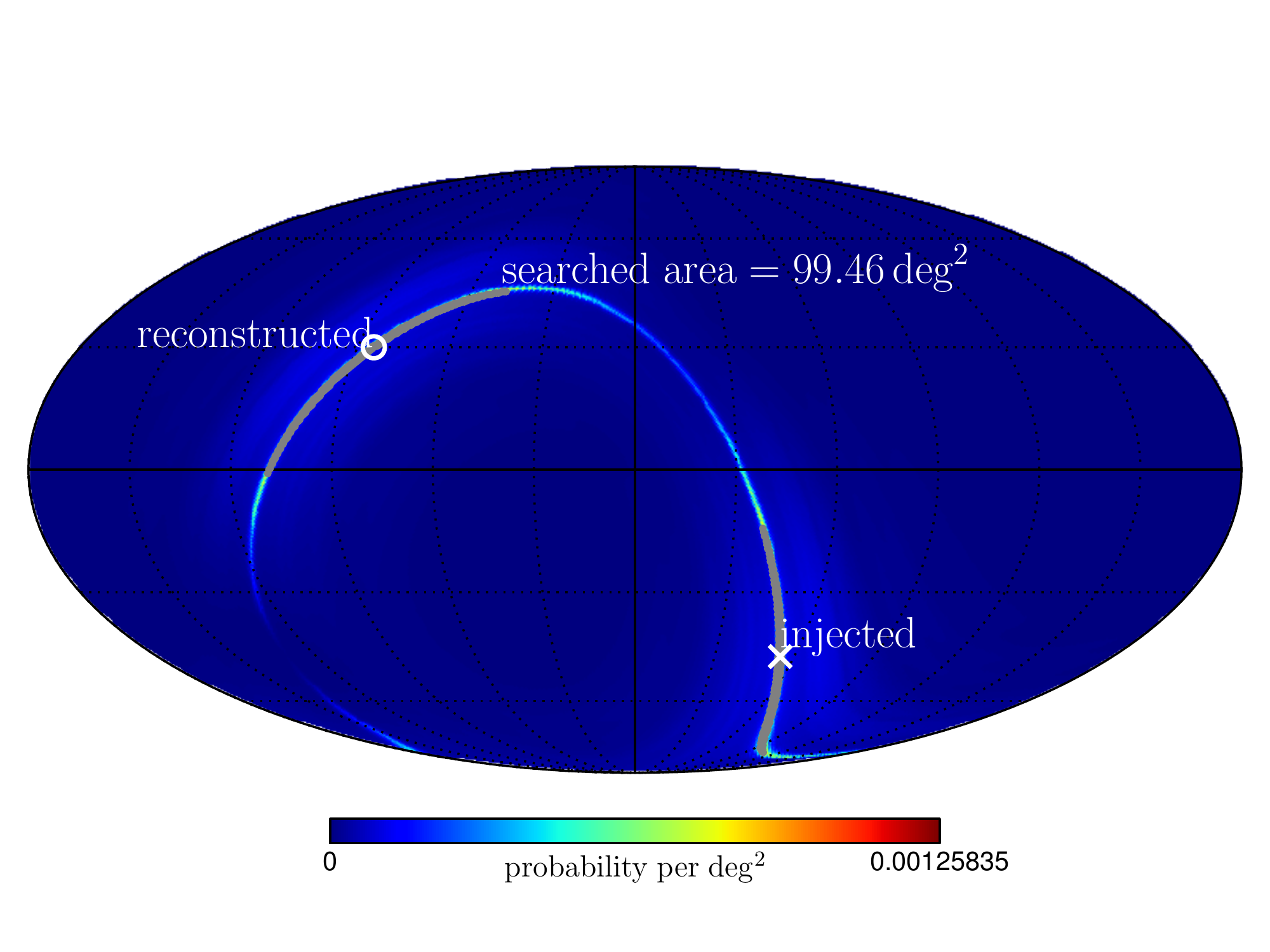} \\
			(c) searched area \\
		\end{center}
	\end{minipage}
	\caption{Mollweide projections demonstrating the two basic 
statistics used to quantify sky localization: angular offsets and
 searched area. 
(a) The entire posterior with the 
injected and reconstructed locations marked. In this example, 
the difference between time-of-arrival in the two LIGO detectors 
(Livingston, LA and Hanford, WA) was 3 ms. (b) The 
\emph{angular offset} is defined as the smallest angle swept between 
the injected location and the reconstructed location (maximum 
a posteriori), shown here as a geodesic. (c) The 
\emph{searched area} is defined as the area assigned a probability 
greater than or equal to the probability assigned to the 
injected location. In the figure this is the shaded area. 
}
	\label{figure:pedagogy}
\end{figure}

Before we investigate the posterior distributions produced 
by \cWB\ and \LIB, we should understand typical features of 
gravitational-wave localization. The majority of sky 
localization comes from time-of-flight measurements between 
distantly located detectors. A single detector is sensitive 
to nearly the entire sky, so it cannot localize sources well. 
However, because gravitational waves travel at the speed of 
light, the difference in the times of arrival at spatially 
separated detectors allow us to triangulate the source location 
on the sky. Figure \ref{figure:pedagogy} \Rev{sketches a skymap generated} with 
two detectors, in which the locus of source positions 
consistent with the observed time-of-flight between detectors 
is a ring. For three detectors, the locus is reduced to two points, 
and so on. \footnote{Section \ref{appendix:sample posteriors} shows a few
sample posteriors produced by each algorithm with our data set.}
This triangulation can be thought of as producing 
a likelihood in the Bayesian sense. Modulation around the 
ring is achieved through knowledge of the antenna patterns and
 an assumption about the distribution of signal amplitudes. 
The detector network is simply more sensitive in some 
directions than others, which means it can detect more signals 
from certain directions. This is included in the effective 
prior for \cWB\ and the prior on template amplitude for \LIB. 
Figure \ref{figure:pedagogy}(a) demonstrates this with clear 
hot spots on the ring, although the entire ring is visible.

Table \ref{table:sample size} 
shows the number of detected events. We injected roughly 100,000 
events for each morphology in each detector network. The low number 
of recovered signals reflects the uniform-in-volume 
distribution of the signals, which causes the majority of 
signals to be very distant and too quiet to be detectable.
Both \cWB\ and \LIB\ were run over the same events. We use all 
events detected by \cWB\ for \G, \SG, and \WNB\ morphologies 
in both years. However, due to the large sample of detected 
\BBH\ injections, we randomly select 500 events 
to process with \LIB. This gives us an accurate 
representation of \LIB's performance with slightly larger errors.

\begin{table}
	\caption{Sample sizes of detected events by morphology and year.}
	\begin{center}
        \begin{tabular}{c c|c|c|c|c}
                \hline
                \hline
                \multirow{2}{*}{year} & \multirow{2}{*}{algorithm} & \multicolumn{4}{c}{morphology} \\
                \cline{3-6}
                \multirow{2}{*}{}     & \multirow{2}{*}{} & G & SG & WNB & BBH \\
                \hline
                \hline
                \multirow{2}{*}{2015} & \cWB              & 256 & 1112 & 769 & 2488 \\
		\multirow{2}{*}{}     & \LIB              & 256 & 1112 & 769 & 500 \\        
		\hline
                \multirow{2}{*}{2016} & \cWB              & 417 & 677 & 853 & 6394 \\
		\multirow{2}{*}{}     & \LIB              & 416 & 664 & 851 & 498 \\
                \hline
        \end{tabular}
	\end{center}
	\label{table:sample size}
\end{table}

\subsection{Angular offset}\label{section:cos_dtheta}

Perhaps the simplest measure of localization is the angle 
between the \emph{maximum a posteriori} and the source's 
position $(\delta\theta)$. Figure \ref{figure:pedagogy}(b) shows 
this for a cartoon posterior, and the line represents 
a geodesic connecting the injected and reconstructed 
locations. In this example, the angle is large because the 
reconstructed location is placed on the wrong side of the ring. 

\begin{figure*}[hb]
  \begin{minipage}{0.5\textwidth}
    \begin{center}
      \includegraphics[width=1.0\textwidth]{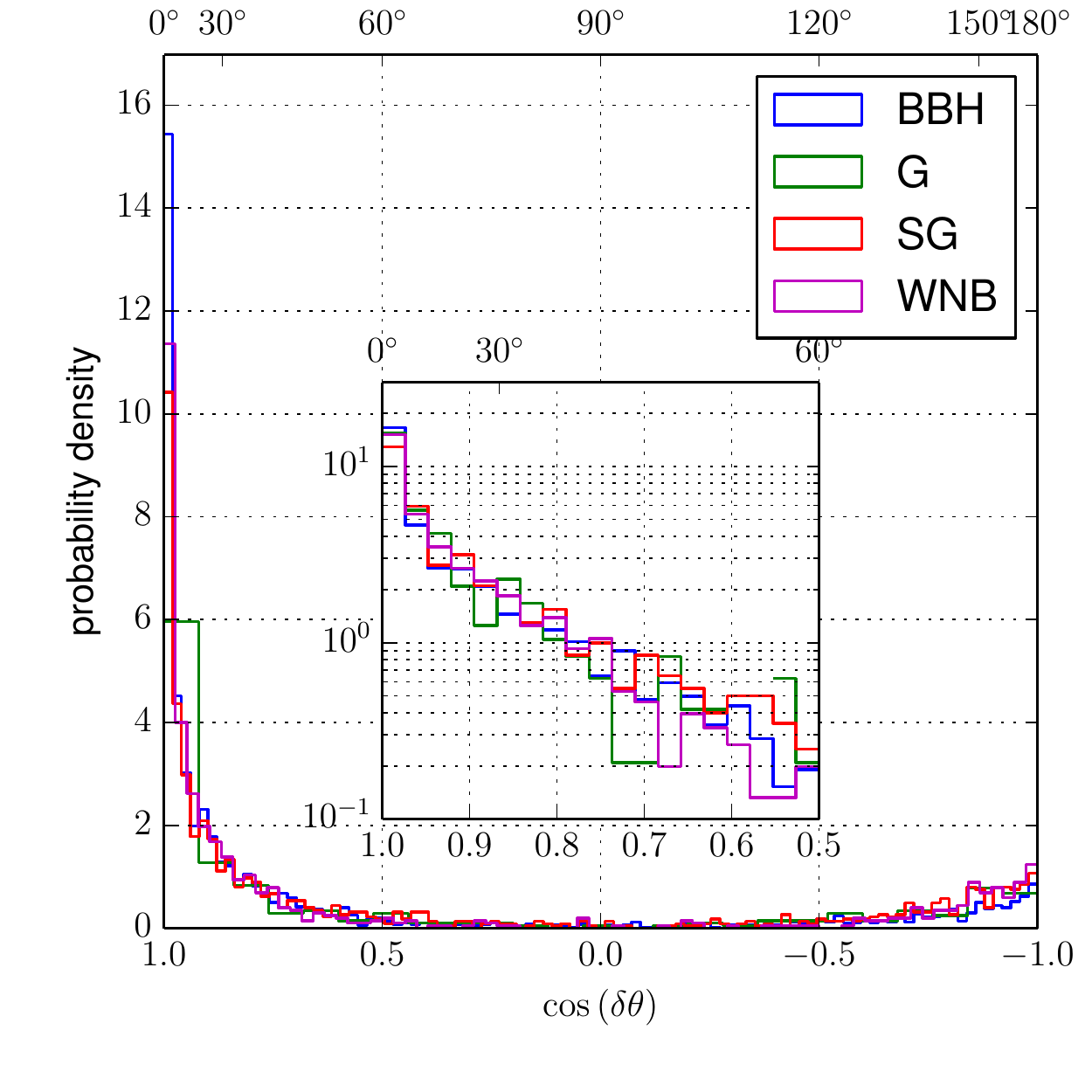} \\
       (a) \cWB\ HL 2015 \\
      \includegraphics[width=1.0\textwidth]{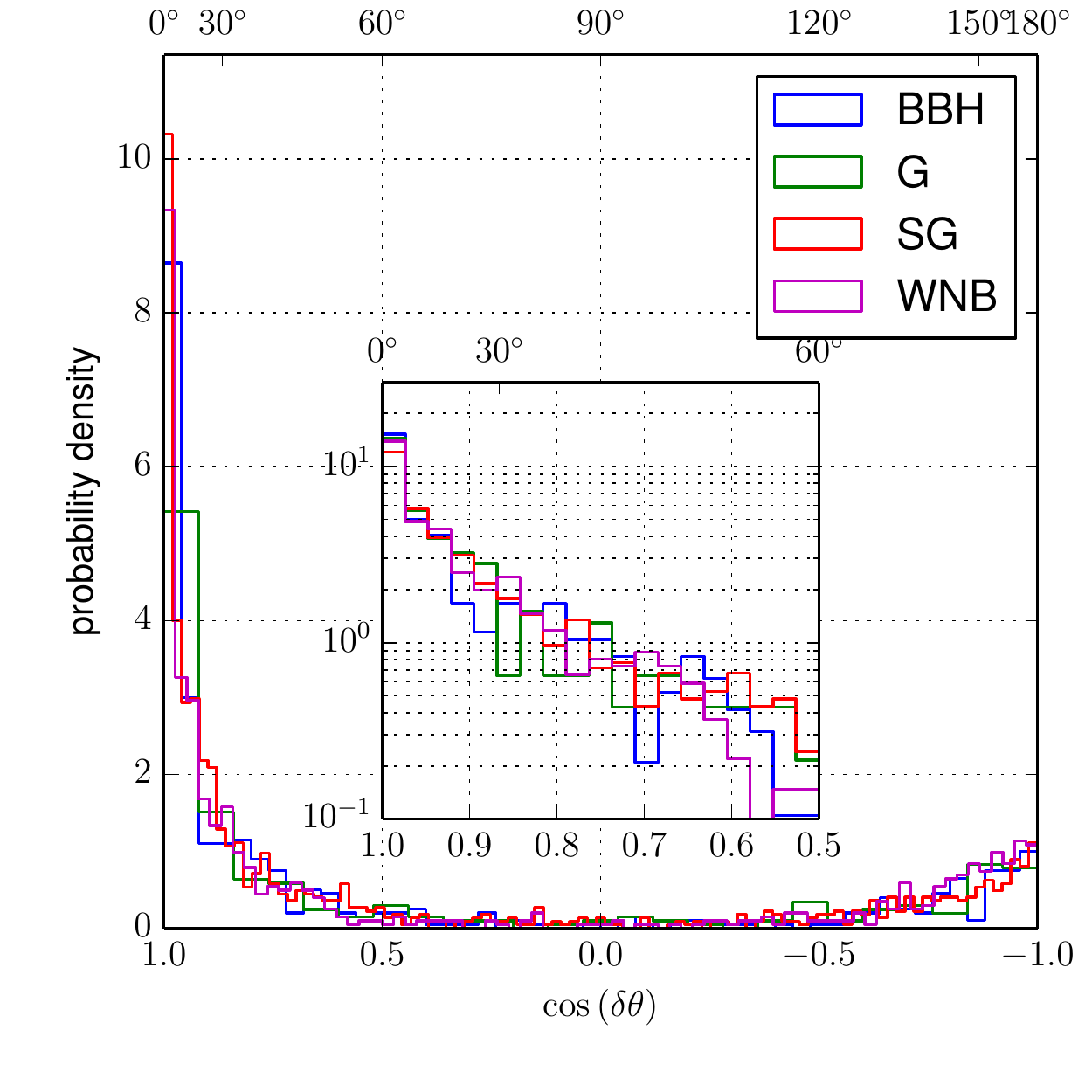} \
       (c) \LIB\ HL 2015 \\
    \end{center}
  \end{minipage}
  \begin{minipage}{0.5\textwidth}
    \begin{center}
      \includegraphics[width=1.0\textwidth]{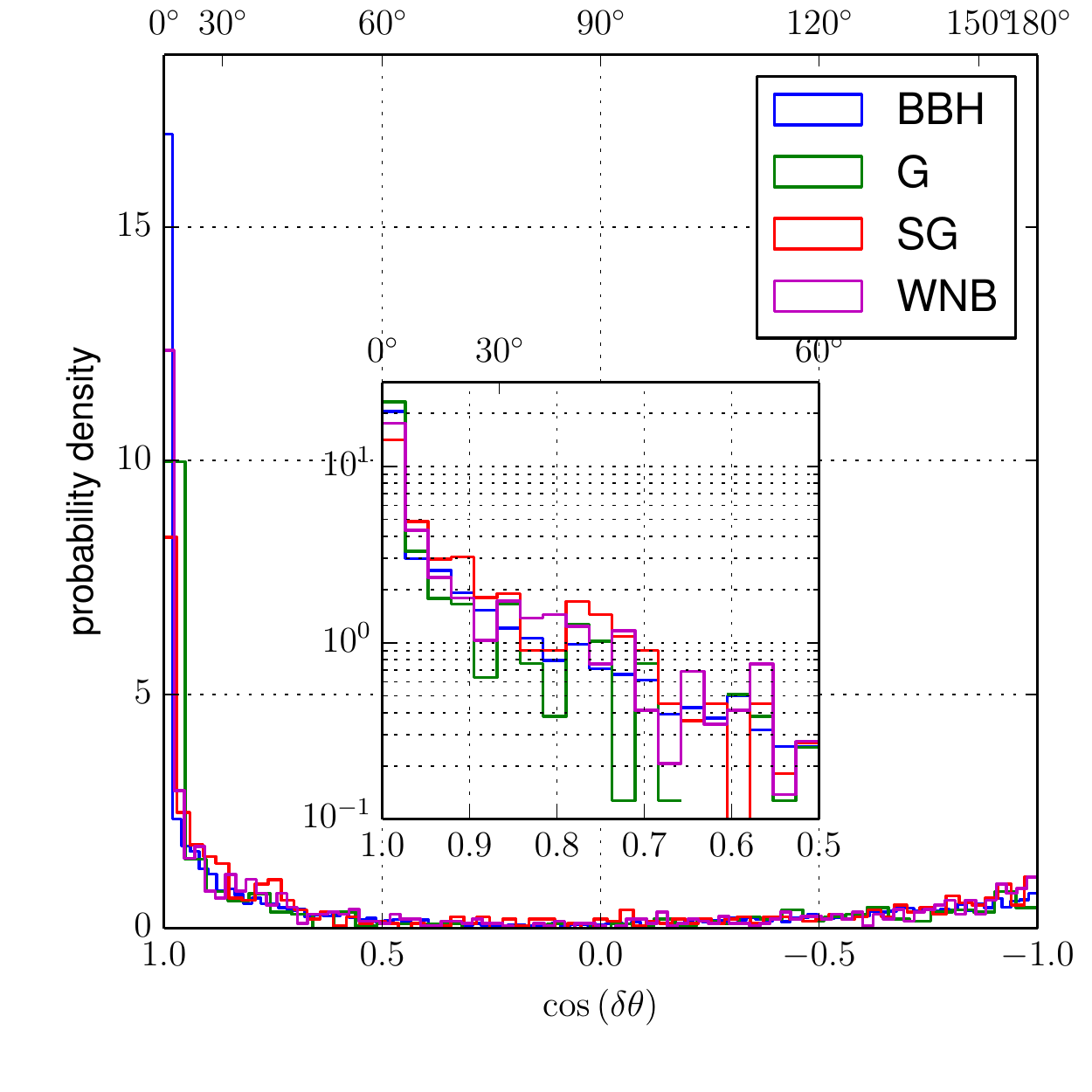} \\
       (b) \cWB HLV 2016 \\
      \includegraphics[width=1.0\textwidth]{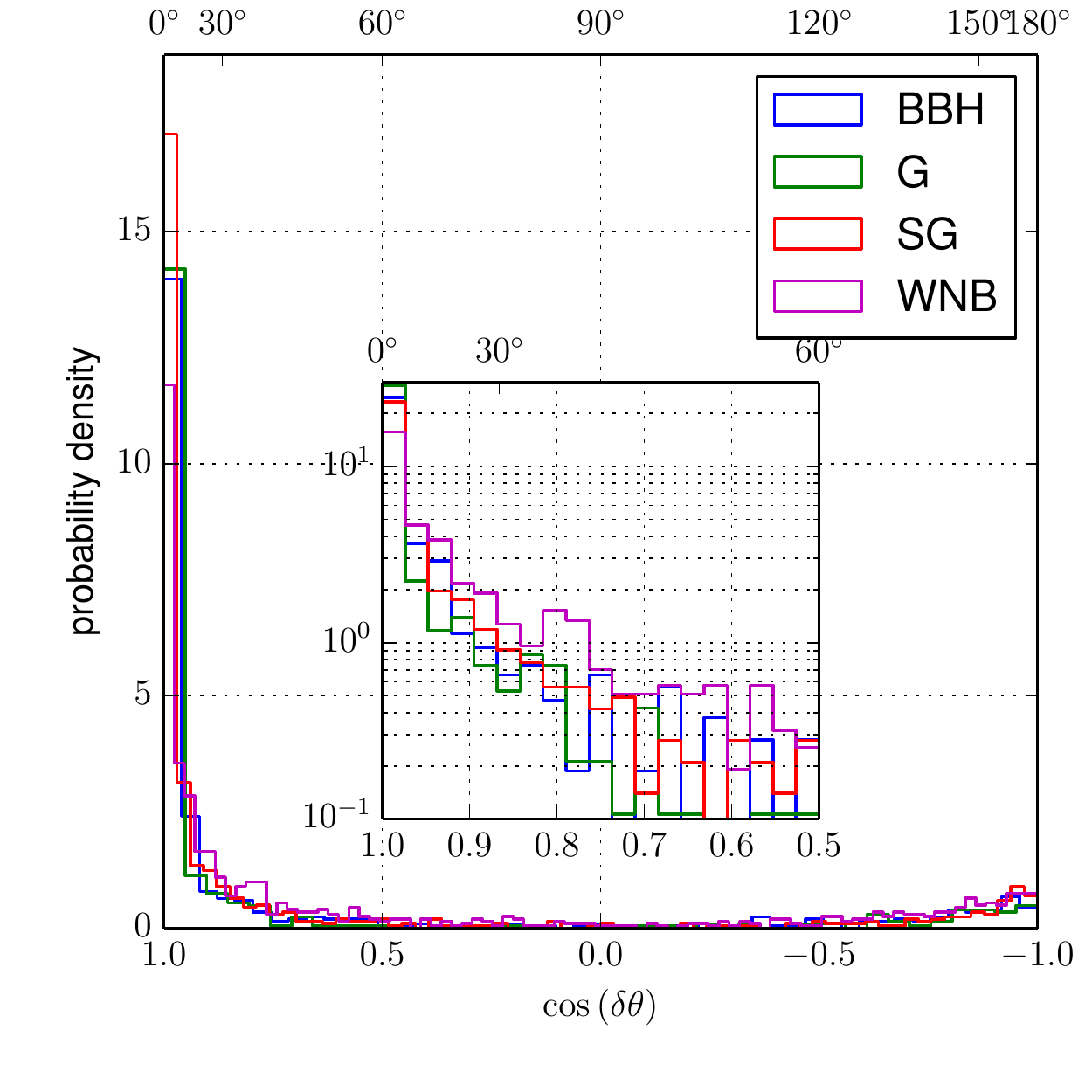} \\
       (c) \LIB\ HLV 2016 \\      
    \end{center}
  \end{minipage}
  \caption{Normalized histograms of $\cos(\delta\theta)$, 
where $\delta\theta$ is the angle between the injected 
location and the maximum of the posterior. (a,c) 2015 detector 
network. (b,d) 2016 detector network. (a,b) \cWB. (c,d) \LIB.}
  \label{figure:ang offsets}
\end{figure*}

Figure \ref{figure:ang offsets} shows the observed 
distributions of this statistic. We use $\cos(\delta\theta)$ to 
highlight grouping around $\delta\theta=0^\circ,\,180^\circ$ 
corresponding to the true and antipodal positions of the source,
 respectively. There is improvement when transitioning from 
the two-detector network to the three-detector network, although 
it is not drastic. This is because Virgo is less sensitive than 
the two LIGO detectors, and many detected events are essentially 
detected by only two detectors.

In the 2015 network, both algorithms place signals close to 
$\cos(\delta\theta)=\pm1$ and produce a desert in between. 
This symmetry is due to the nearly aligned antenna patterns 
for the two LIGO detectors, which makes the antipode degenerate 
with the correct side of the sky. This symmetry is a generic 
feature of the detector's antenna patterns and also appears 
when the morphology is known a priori [~\cite{Singer:2014}].
In the 2016 network, there is some reduction in the mode near 
$\cos(\delta\theta)=-1$. We note that the peak near 
$\delta\theta\sim0$ is sharper in 2016 than in 2015, however 
the median value of $\delta\theta$ is actually larger in 2016 
for \cWB\ ($\sim33^\circ$ compared to $23^\circ$ for \WNB). 
This may be due to the regulators (see Section 
\ref{section:systematics}).
In the three-detector network, there is a degeneracy in the posteriors associated with 
reflections about the plane defined by the three detectors 
[~\cite{Aasi:2013,Fairhurst:2009}]. This degeneracy may be 
responsible for some of the antipodal population as well. This 
can be seen in the \LIB\ distributions, which show the degeneracy 
but also show a decrease in the median $\delta\theta$ in the 
2016 network, as expected.

We also note that the general features of source localization 
by both \cWB\ and \LIB\ do not depend strongly on the signal 
morphologies. This is significant, as it suggests that the 
same localization algorithm can be used to construct posteriors 
for generic bursts without waveform-specific biases. 

Typically, we expect $\cos(\delta\theta)\rightarrow1$ in the 
limit of high SNR. This holds for the three-detector network, in 
which the signal is triangulated to the true location, and
 degeneracy from reflections defined by the plane of the 
detectors are broken by the antenna patterns. However, this is 
not the case in the two-detector network. This is because, 
even in the limit of high SNR, the two-detector network cannot 
break the ring-degeneracy in the posterior. If the waveform is
 not known a priori, then it must be reconstructed. For the 
two-detector network, different points along the triangulation
 ring correspond to different reconstructed waveforms. However, 
they all reproduce the data equally well and the algorithm 
does not prefer one location on the ring over another. In fact, 
only the width of the ring decreases as SNR increases. 

Tables \ref{table:cWB statistics} and \ref{table:LIB statistics} 
give cumulative fractions of events with $\delta\theta$
 less than a few exemplar values.

\subsection{Searched area}\label{section:searched area}

Another measure of source localization is the \emph{searched area}. 
We define this as the area on the sky assigned a probability 
greater than or equal to the probability assigned to the 
injected location [~\cite{Abadie:2012c}]. If a follow-up 
algorithm sorts through an list of pixels ordered by 
probability, this approximates the amount of area imaged before 
finding the injection. Importantly, the searched area does 
not account for spatially separated support in the posterior 
distribution. Figure \ref{figure:pedagogy}(c) demonstrates 
this statistic as the shaded area. 
We can estimate the fraction of events for which electromagnetic
counterparts will be observed within a given area with a cumulative
distribution over the searched area.
\footnote{Another possible observing plan would be to set a 
confidence regions, say 50\%, rather than a fixed area. 
We could then determine the fraction of events with 50\% 
confidence regions containing less than a given area. 
We avoid this statement here because the posteriors produced 
by \cWB\ and sometimes \LIB\ are poorly calibrated (see 
section \ref{section:systematics}).}

\begin{figure*}
  \begin{minipage}{0.5\textwidth}
    \begin{center}
      \includegraphics[width=1.0\textwidth]{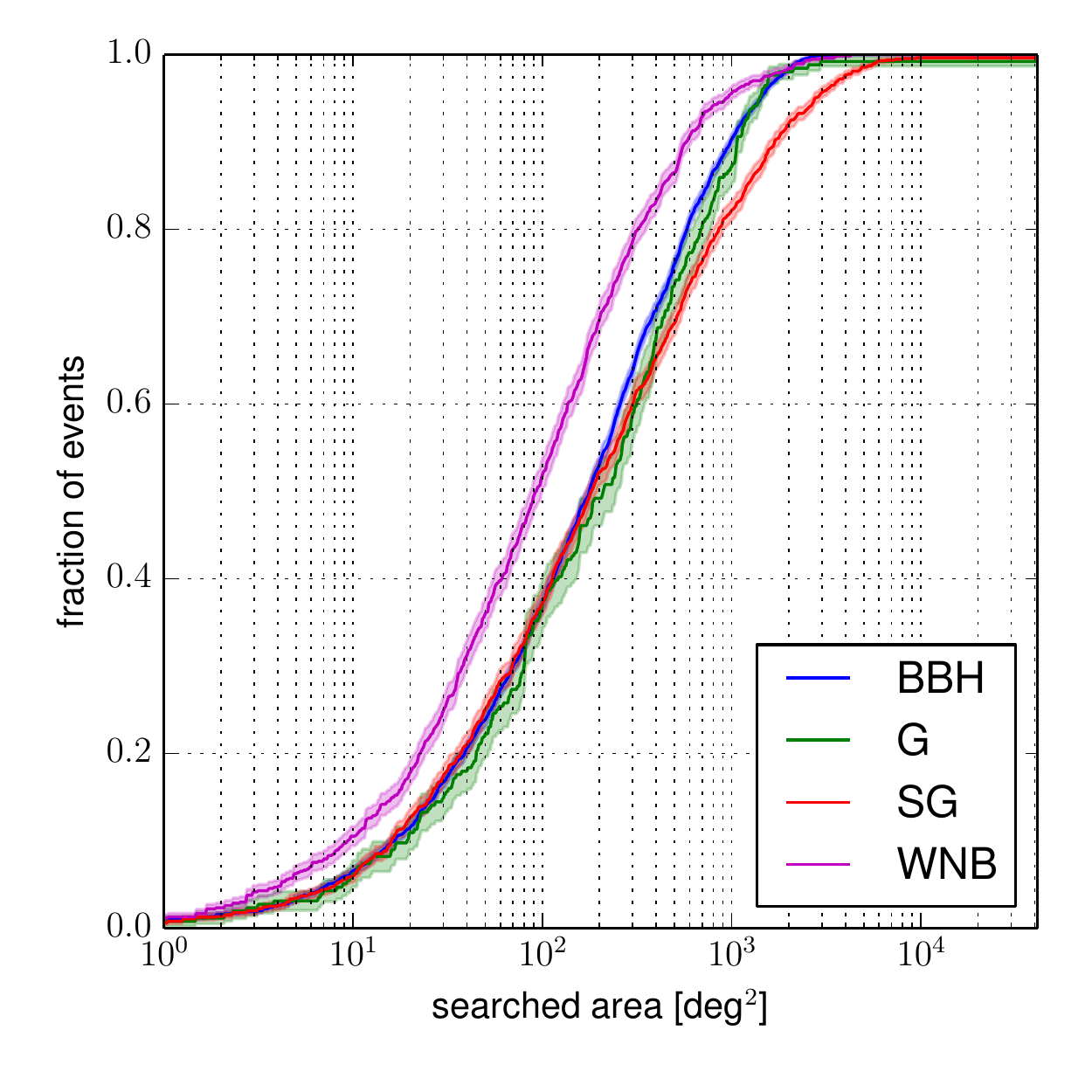} \\
       (a) \cWB\ HL 2015 \\
      \includegraphics[width=1.0\textwidth]{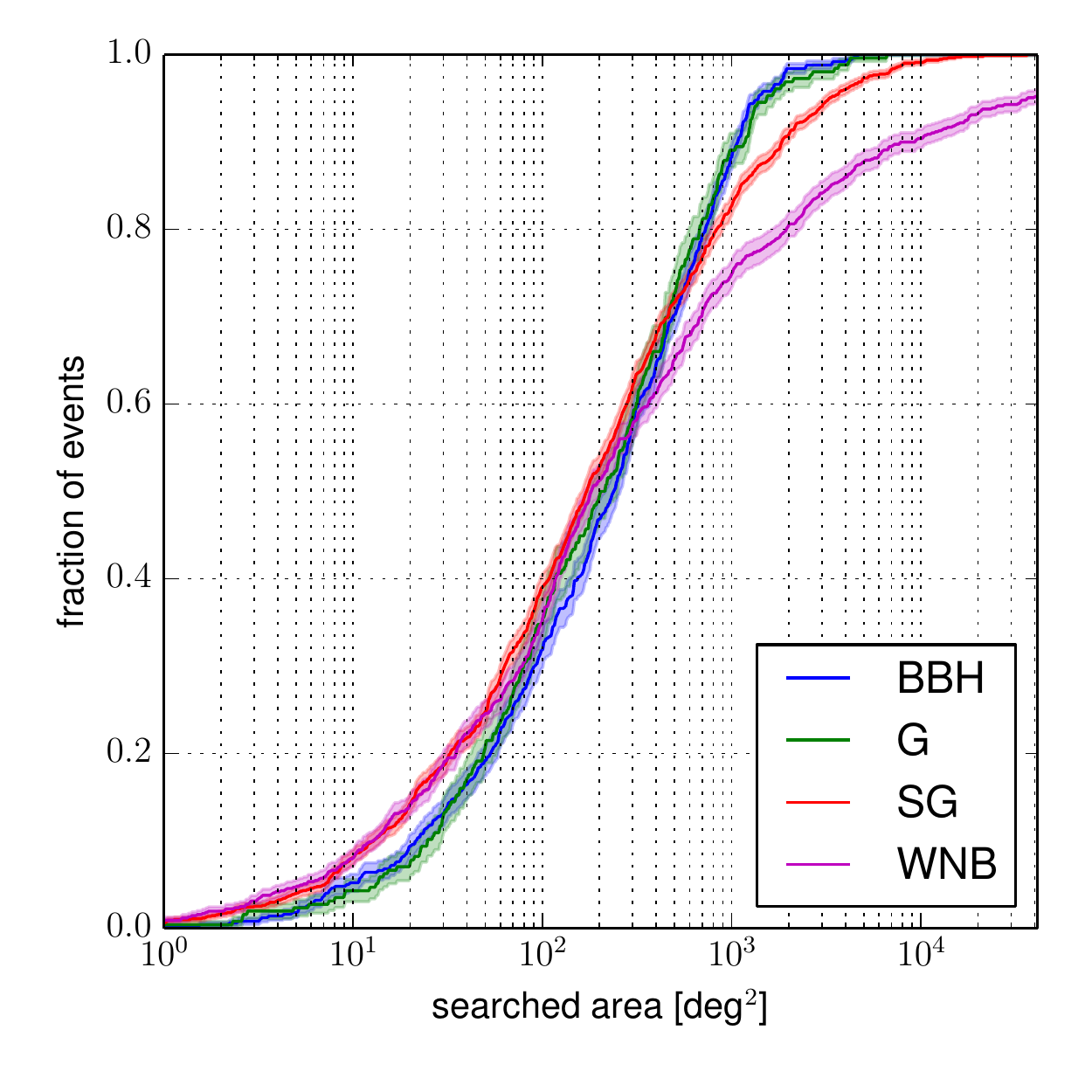} \\
       (c) \LIB\ HL 2015 \\
    \end{center}
  \end{minipage}
  \begin{minipage}{0.5\textwidth}
    \begin{center}
      \includegraphics[width=1.0\textwidth]{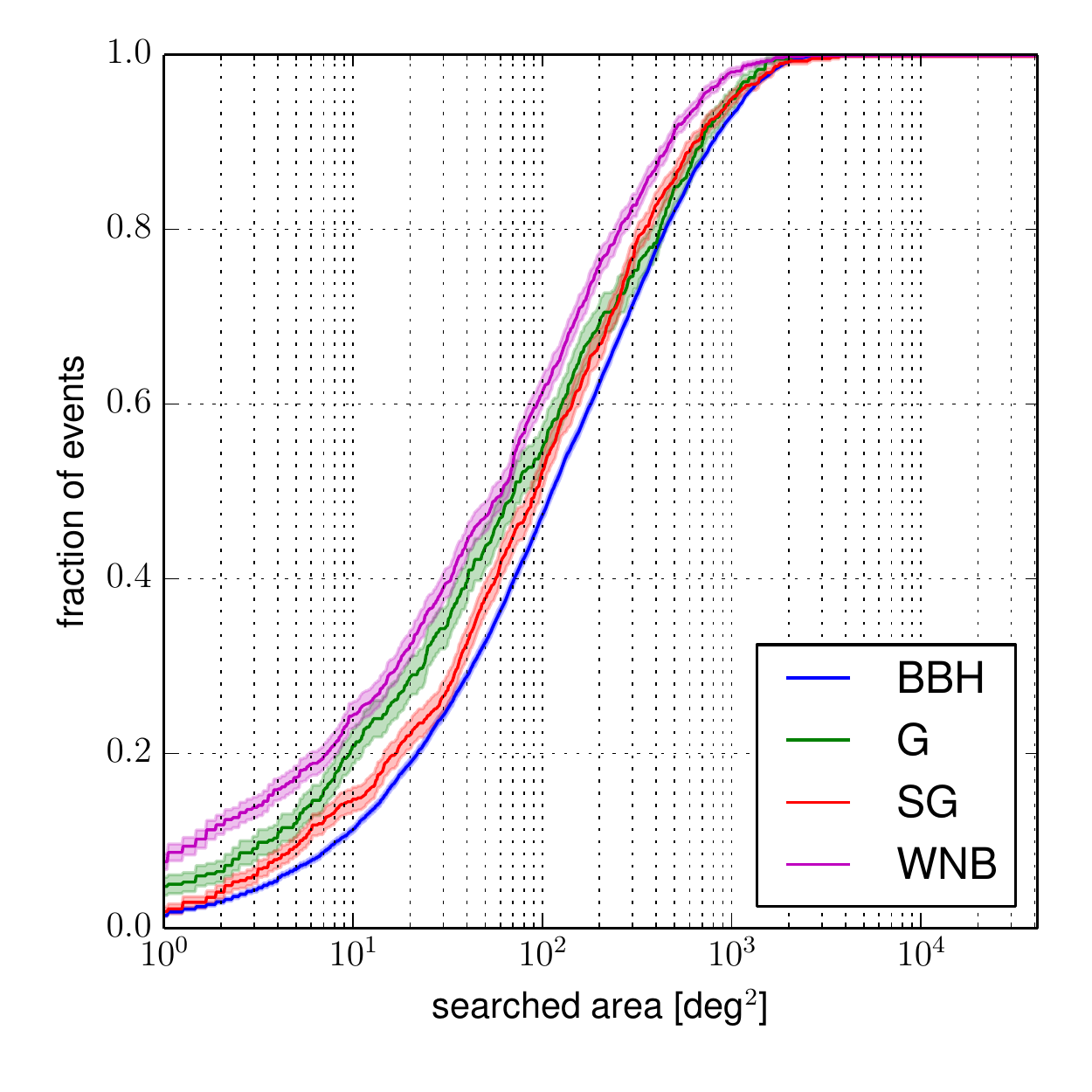} \\
       (b) \cWB\ HLV 2016 \\
      \includegraphics[width=1.0\textwidth]{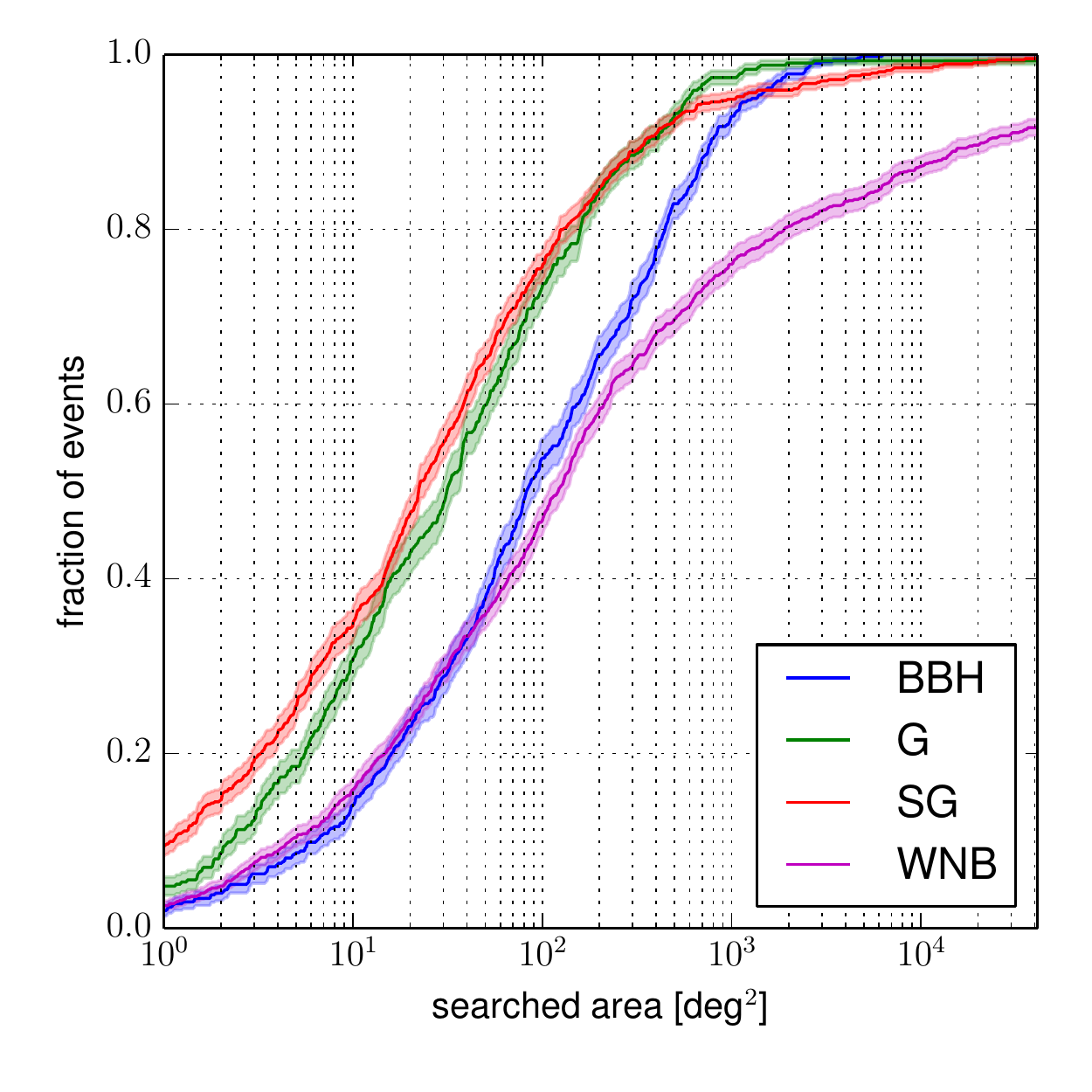} \\
       (d) \LIB\ HLV 2016 \\
    \end{center}
  \end{minipage}
  \caption{Cumulative histograms of searched areas for (a,c)
 2015 HL and (b,d) 2016 HLV; (a,b) \cWB\ and (c,d) \LIB. 
Shaded regions correspond to 68\% confidence intervals.}
  \label{figure:searched areas}
\end{figure*}

Figure \ref{figure:searched areas} shows cumulative 
distributions of observed searched areas for all morphologies 
considered. Importantly, we see that the searched area improves 
when moving from the two-detector network in 2015 to the 
three-detector network in 2016, even though the Virgo noise 
curve is nearly twice as high as the LIGO curves in 2016. This 
can be attributed to a more informative likelihood, and is true 
for both algorithms. 

We also see that \LIB\ performs much worse than \cWB\ for 
\WNB\ signals. This is due to template mis-match within \LIB, 
which attempts to model all signals with a single \SG\ template. In
 fact, \LIB\ assigns a few \WNB\ signals searched areas 
equal to the entire sky. This is because the random \WNB\ 
waveform matches so poorly with the \SG\ template that \LIB\ 
does not detect the signal. 

In 2015, both algorithms perform nearly identically for both
 \SG\ and \G\ waveforms. This reflects the fact that we 
cannot construct posteriors for burst signals more accurately 
than a timing ring modulated by the antenna patterns. \LIB\ 
outperforms \cWB\ for the smallest searched areas and for \SG\ 
signals, as expected given the waveform basis used in \LIB. 
However, the fact that both algorithms agree over a wide range 
of searched areas suggests there is only minimal improvement 
possible with knowledge of the actual signal morphology.

In 2016, \LIB\ localizes \G\ and \SG\ signals 
better than \cWB. This is because \LIB\ uses \SG\ templates 
to recover signals, while \cWB\ does not. Therefore, knowledge 
of the correct signal morphology can significantly improve 
localization in three-detector networks. However the performance 
between the algorithms is more comparable for \BBH\ signals.

We also note that \WNB\ are localized consistently better 
than \SG, \G, or \BBH\ with \cWB. This should not be surprising. 
A simple Fisher matrix computation like those in 
[~\cite{Fairhurst:2009,Fairhurst:2011}] shows that, for 
\SG\ signals, the expected errors in time-of-flight between 
detectors should scale as

\begin{equation}
\sigma_t^2 \sim \frac{1}{\rho^2}\left( f_o^2 + \tau^{-2} \right)^{-1} .
\end{equation}

$\tau^{-1}$ is related to the signal's bandwidth, and 
therefore we expect high frequency, high bandwidth signals 
to have the smallest timing errors. From our injections, 
\WNB\ signals will typically have higher frequencies than \G\ 
and larger bandwidths than \SG. This means they will have 
smaller timing errors and narrower triangulation rings.

Tables \ref{table:cWB statistics} and \ref{table:LIB statistics} 
give cumulative fractions of events with searched area less 
than a few exemplar values.
Unlike $\cos(\delta\theta)$, we see a strong decrease in the
 searched area in the limit of high SNR in both 2015 and 2016.
This is expected from simple triangulation, and corresponds
to a narrowing of the triangulation ring in the two-detector
case rather than the removal of the ring.

\subsection{Extent of the posterior's support}

While the searched area and $\delta\theta$ are good indicators 
of the localization, they do not describe the entire posterior.
For example, both the searched area and $\delta\theta$ may be 
small, suggesting a compact posterior distribution. However, 
the little area that is included may be scattered across 
the sky, with a small $\delta\theta$ merely fortuitous. This
 could correspond to a very narrow ring in the two-detector 
case, with the reconstructed location placed next to the 
injected location by chance. 

\begin{figure*}
  \begin{minipage}{0.5\textwidth}
    \begin{center}
      \includegraphics[width=1.0\textwidth]{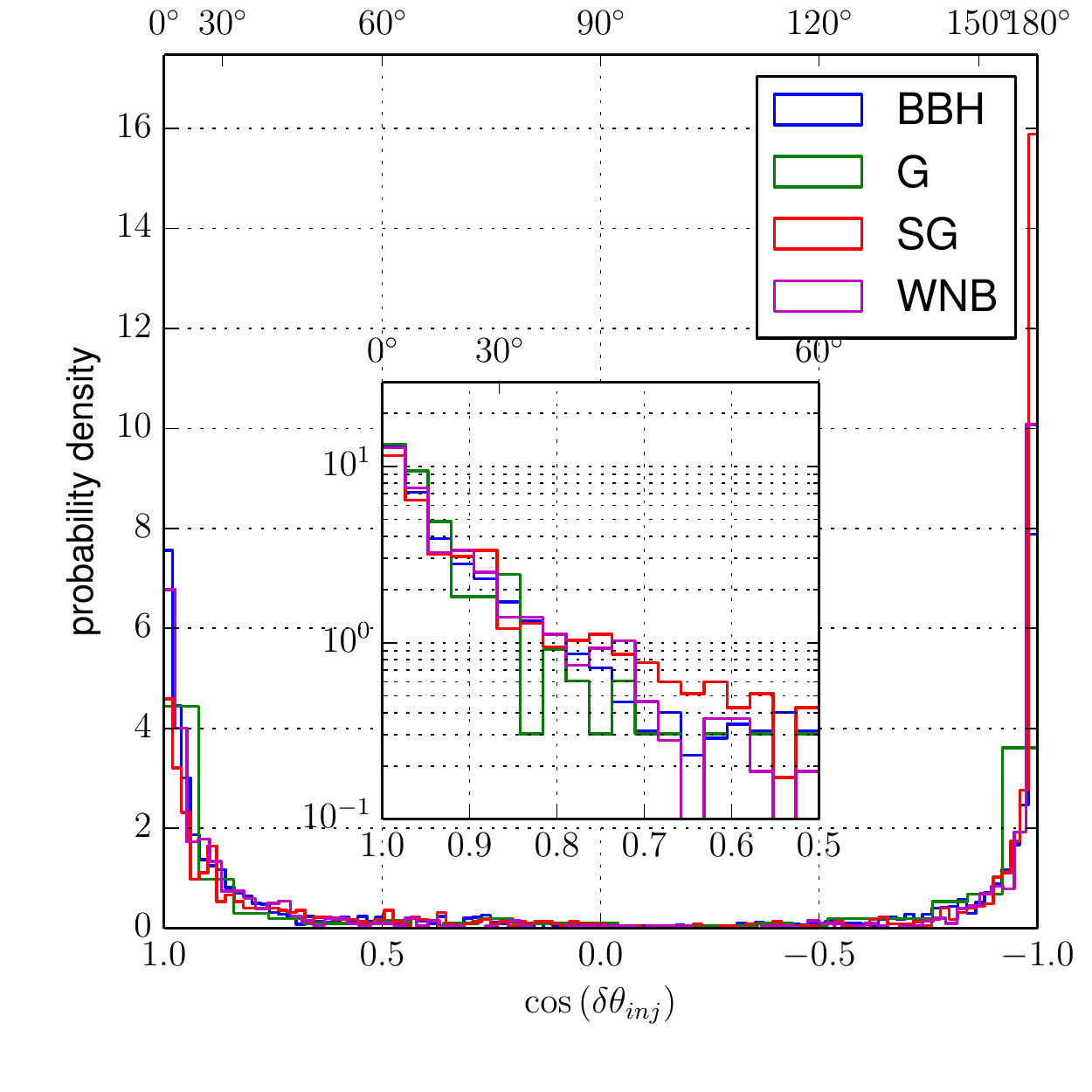} \\
       (a) \cWB\ HL 2015 \\
      \includegraphics[width=1.0\textwidth]{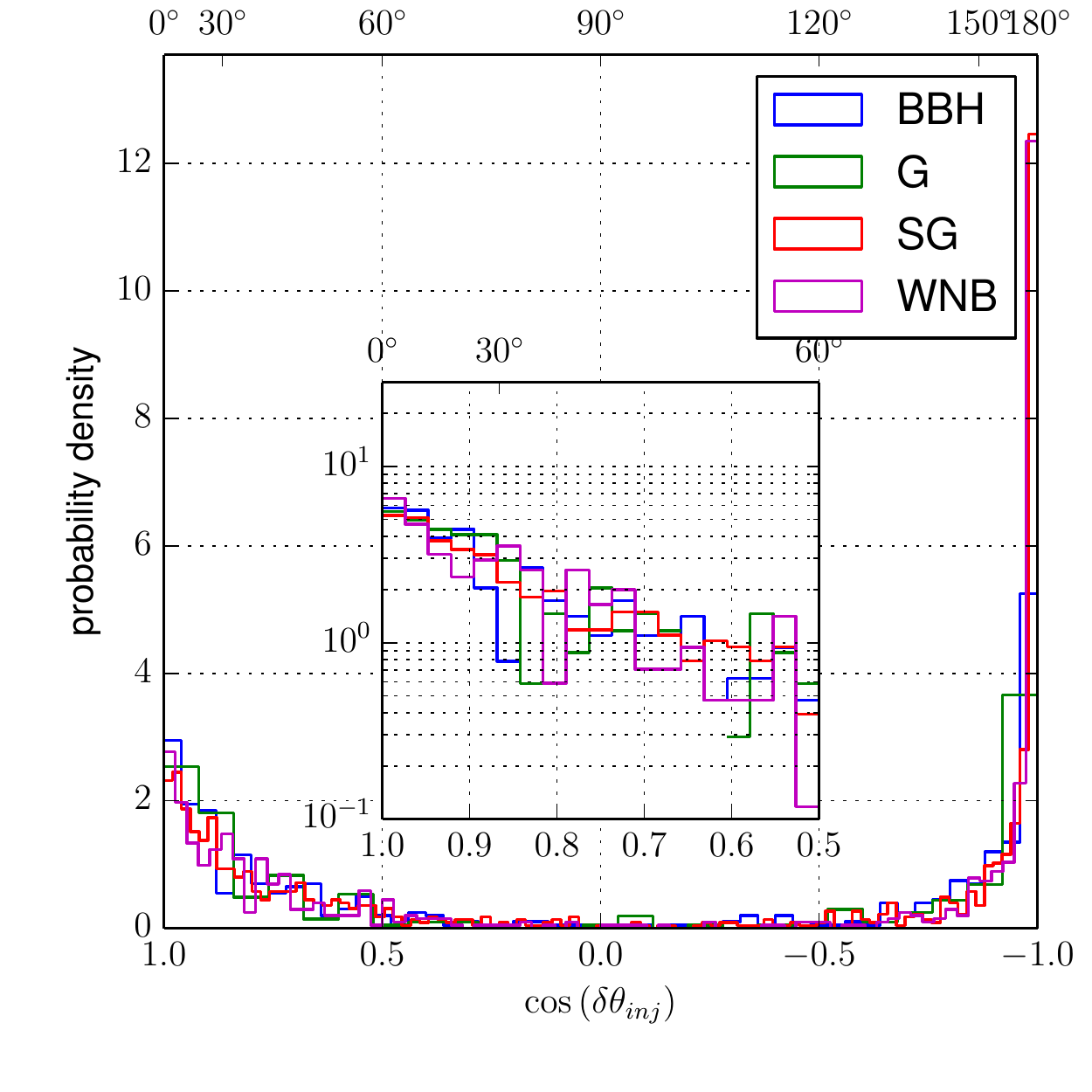} \\
       (c) \LIB\ HL 2015 \\
    \end{center}
  \end{minipage}
  \begin{minipage}{0.5\textwidth}
    \begin{center}
      \includegraphics[width=1.0\textwidth]{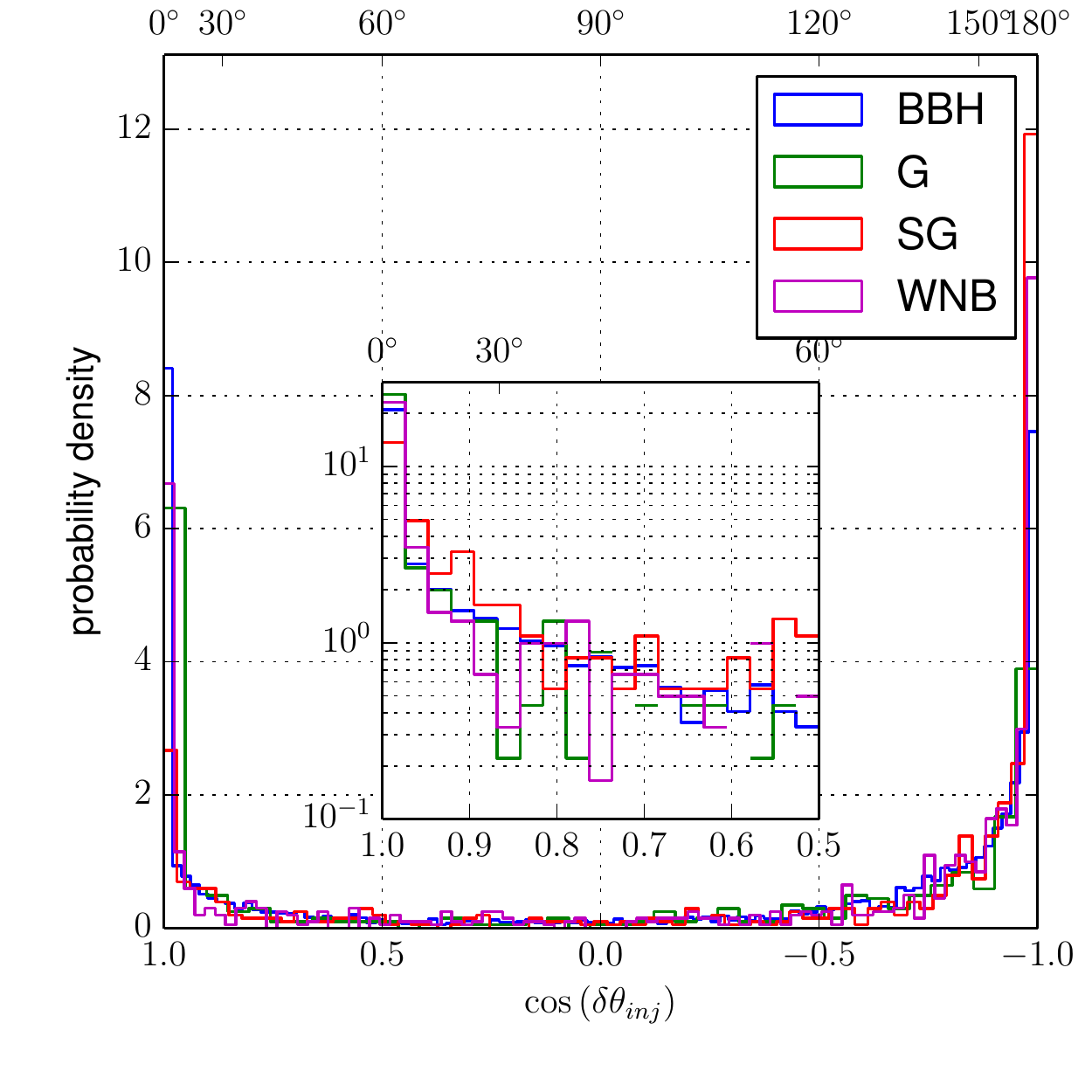} \\
       (b) \cWB\ HLV 2016 \\
      \includegraphics[width=1.0\textwidth]{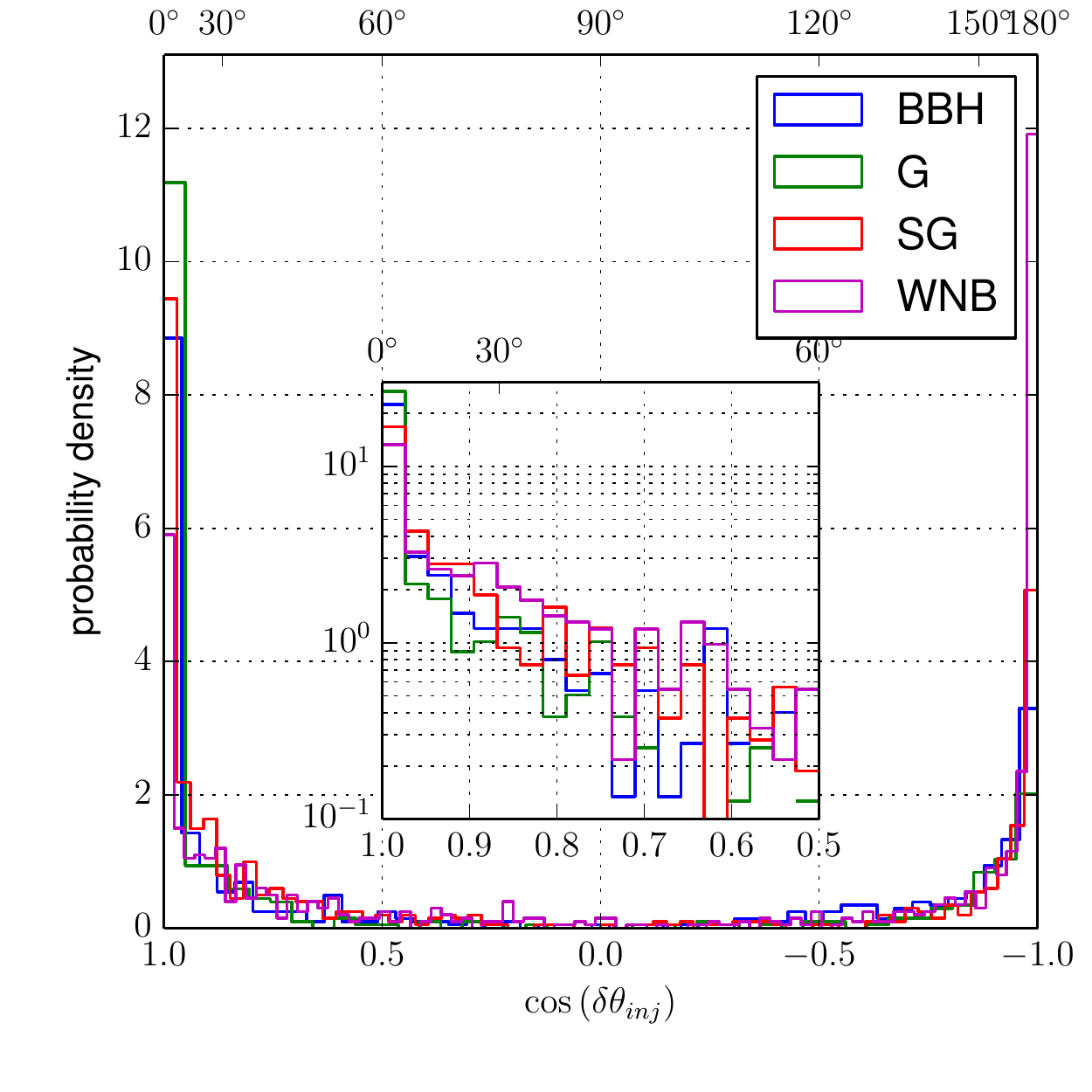} \\
       (d) \LIB\ HLV 2016 \\
    \end{center}
  \end{minipage}
  \caption{Normalized histograms of the largest angle between 
the source's true location and any point within the searched 
area. (a,c) 2015 network. (b,d) 2016 network. (a,b) \cWB.
 (c,d) \LIB. 
}
  \label{figure:largest distance between points within searched area}
\end{figure*}

To diagnose the prevalence of such cases, we plot the maximum 
angular distances from the injection's source to any point 
in the searched area ($\delta\theta_{inj}$) in Figure 
\ref{figure:largest distance between points within searched area}. 
If an electromagnetic follow-up is carried out systematically 
over this area, this estimates the separation between points in
the region searched before imaging the source. In particular, 
we should be able to determine whether the posterior has 
support at antipodal points in the sky, which are difficult 
to observe with a single telescope. 

Figure \ref{figure:largest distance between points within searched area} 
shows that there is support near the antipode for a large 
fraction of events in 2015. This indicates that we will find 
support all along the degenerate ring modulated by the 
antenna patterns, which happens to be on the other side of 
the sky a significant fraction of the time. 
The lobe near $\delta\theta_{inj}\sim0^\circ$ corresponds to
 small searched areas, in which the injection was found quickly 
before the searched area included points from the antipodal 
antenna pattern maximum.

The three-detector network shows similar structure. We again see
 a population of events with small searched areas, with 
$\delta\theta_{inj}\sim0^\circ$. Somewhat surprisingly, we see a 
large lobe near $\delta\theta_{inj}\sim180^\circ$. This is likely 
due to a combination of Virgo's higher noise curve and the 
reflection degeneracy visible in the $\delta\theta$ 
distribution as well (\rev{S}ection \ref{section:cos_dtheta}). 
With fewer than \rev{four} detectors, these antipodal degeneracies 
may be unavoidable for un-modeled signals. 

\subsection{Fragmentation}

Furthermore, the largest angle between points in the searched 
area does not tell us about the posterior's shape.
Support could be placed along a large ring or randomly 
scattered in distant parts of the sky. We call this the 
fragmentation of the posterior and attempt to measure it by 
counting the number of disjoint regions within a 
specified area. For example, if the posterior is split between 
a blob and it's antipode, there are two. This is the case for 
the searched area in Figure \ref{figure:pedagogy}(c).

\begin{figure*}
  \begin{minipage}{0.5\textwidth}
    \begin{center}
      \includegraphics[width=1.0\textwidth]{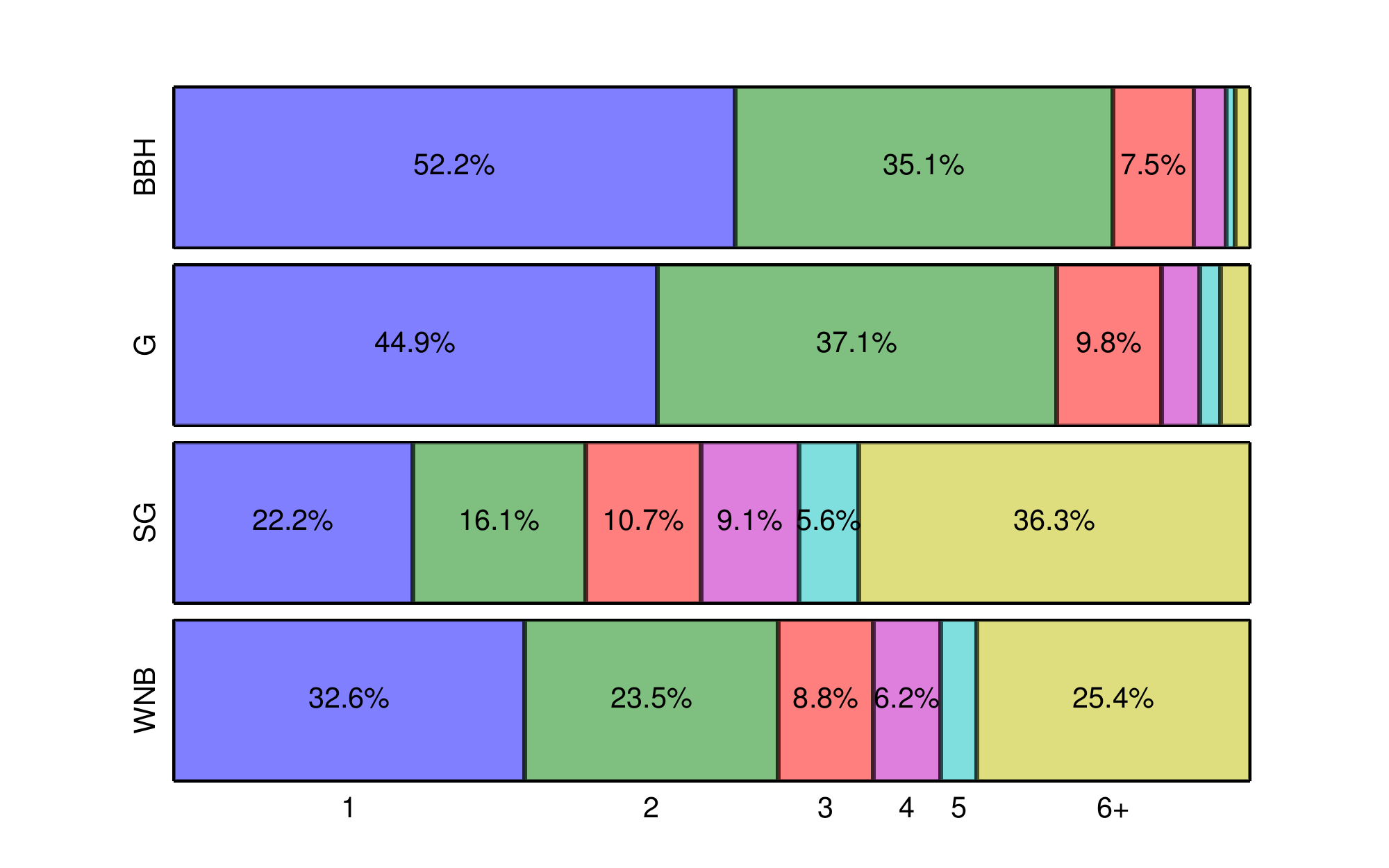} \\
       (a) \cWB\ HL 2015 \\
      \includegraphics[width=1.0\textwidth]{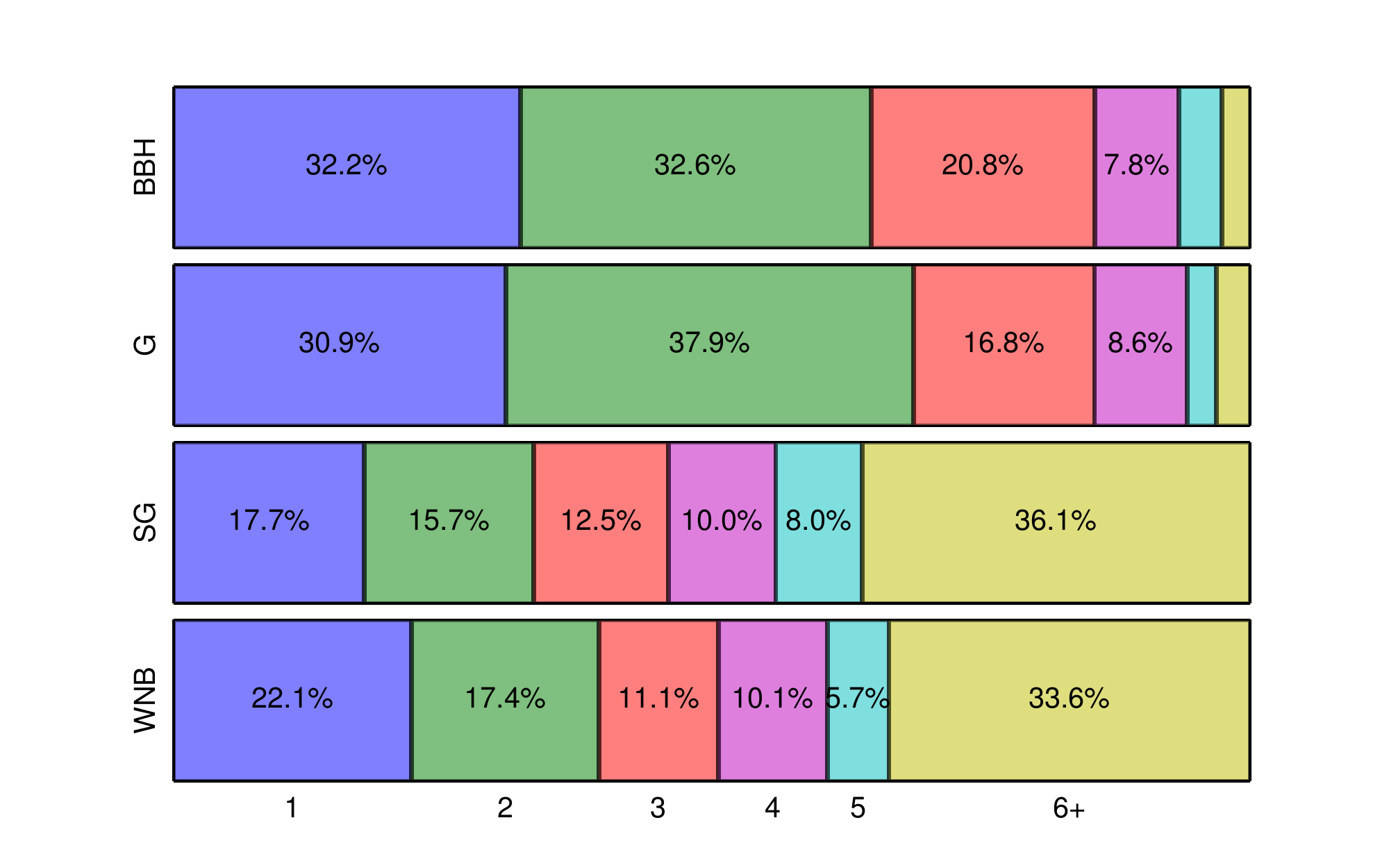} \\
       (c) \LIB\ HL 2015 \\
    \end{center}
  \end{minipage}
  \begin{minipage}{0.5\textwidth}
    \begin{center}
      \includegraphics[width=1.0\textwidth]{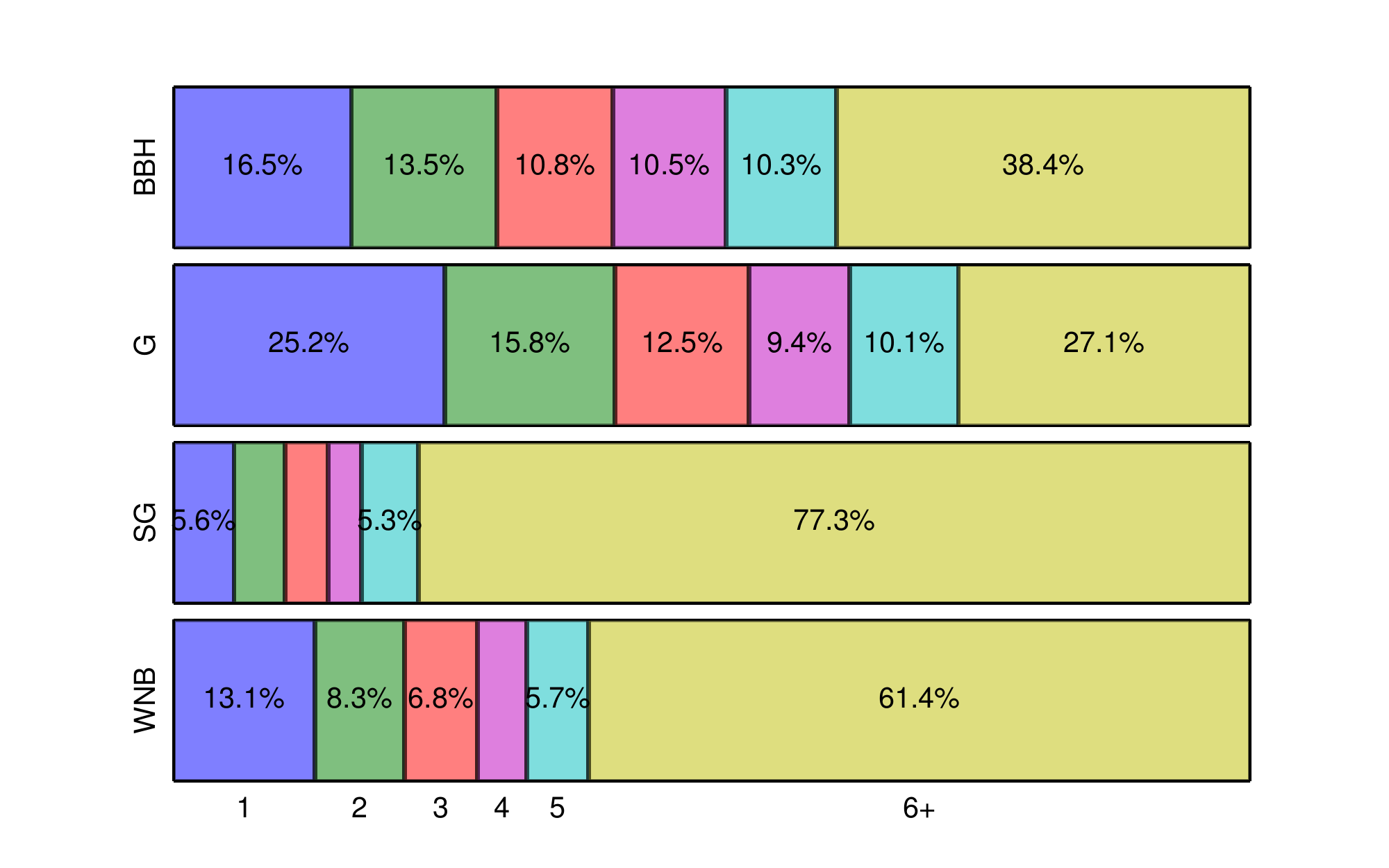} \\
       (b) \cWB\ HLV 2016 \\
      \includegraphics[width=1.0\textwidth]{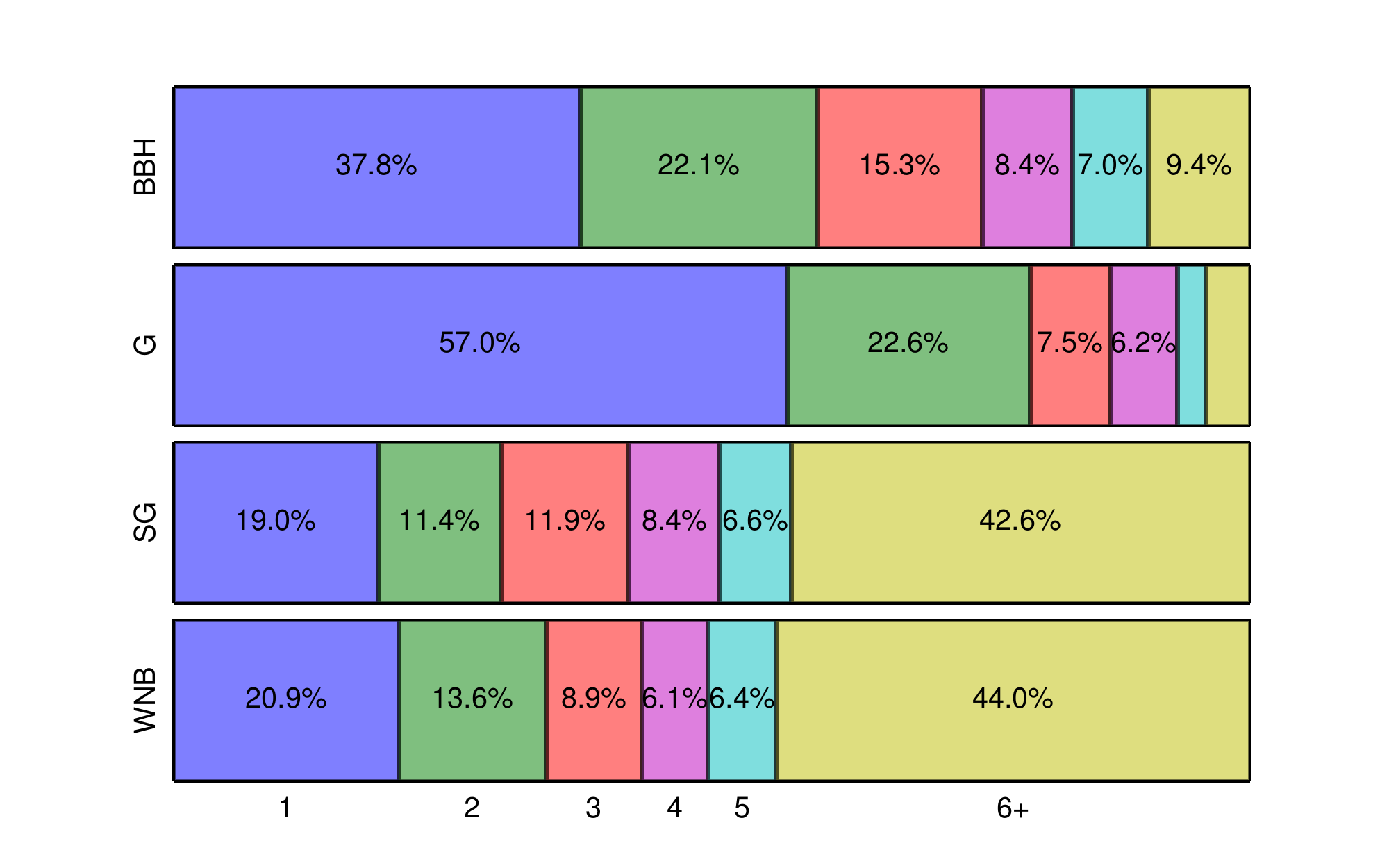} \\
       (d) \LIB\ HLV 2016 \\
    \end{center}
  \end{minipage}
  \caption{Fractions of events binned by the number of disjoint 
regions in the searched area. (a,c) 2015 HL and 
(b,c) 2016 HLV. (a,b) \cWB. (c,d) \LIB. \SG\ signals typically
 have more modes because they support fringe peaks.}
 \label{figure:sky fragmentation}
\end{figure*}

Figure \ref{figure:sky fragmentation} shows histograms of the 
number of simply connected regions within the searched area. 
There is a lot of morphology dependence, but a few trends are 
clear. \G\ and \BBH\ signals typically have fewer simply 
connected regions than \SG\ and \WNB\ signals. For \SG\ signals, 
this is because of their strong central frequency and relatively 
narrow bandwidth. 

The oscillatory waveforms imprinted in the data from each 
detector still match relatively well if they are offset by a 
small number of cycles, which corresponds to a time-of-flight 
error between detectors. This causes \rev{periodic} features 
in the posterior with typical angular scales of 
$\Delta\theta\propto1/f_o$. 
\footnote{\rev{The features are not exactly periodic because the signal may \Rev{vary} significantly over time scales comparable to $1/f_o$ and the antenna patterns may favor only part of the sky.}}
We therefore expect to see parallel 
rings in two-detector posteriors, and a nearly regular lattice 
in the three-detector posteriors. 

We also note that the \WNB\ signals appear to have 
fragmentation somewhere between \G\ and \SG\ signals. This is 
because the \WNB\ signals have wide ranges for their 
bandwidths relative to their central frequencies. When the 
bandwidth is comparable to the central frequency, there are 
no fringe peaks (like a \G\ signal) and when it is narrow 
compared to the central frequency there are fringe peaks 
(like a \SG\ signal). \BBH\ signals act similar to \G\ 
signals because they have relatively broad bandwidths and 
are concentrated at low frequencies.

\subsection{Direct comparison of posteriors}
\label{section:comparison}

While all the previous statistics allow us to investigate the 
relative performance of \cWB\ and \LIB\ through ensemble averages, 
they do not tell us how the sky maps produced by these two methods 
compare on an event-by-event basis.
Comparing localization directly for each event will be important
if all algorithms are used to provide alerts and posteriors to the
astronomical community.

We first investigate the angular 
separations between the maxima a posteriori (MAP).
 There is remarkable agreement in the 2015 data, with 
the median $\delta\theta_{MAP}$ consistently around $8^\circ$. In 
the 2016 data set there is significantly more disagreement, with 
the median $\delta\theta_{MAP}$ as high as $84^\circ$ for some 
morphologies. 

In addition to the angular separation of the maxima a posteriori, 
we examined the intersection and union of sets of pixels selected 
by each algorithm. We compute the ratio of intersection to union ($i/u$)
for pixels selected by the two algorithms. Specifically, we compute $i/u$ 
using 50\% and 90\% frequentist confidence regions. If $i/u$ is near unity, the maps 
are very similar and if $i/u$ is close to zero the maps either select nearly disjoint 
regions or one map includes much more area than the other.
In 2015, the median
$i/u$ is near 0.5 for both confidence levels. However, there is significantly
less overlap in the three-detector network, with $i/u$ between 0.2 and 0.3 even
for the 90\% confidence region.

We also compute the total size of the spatial support in each posterior for 
each algorithm. In both the two-detector and three-detector networks, \LIB\ typically assigns 
non-zero probabilities to many more pixels than \cWB. 
We note that \cWB\ typically assigns a lower confidence to the same 
region surrounding the maximum a posteriori than \LIB. Combining this 
with the knowledge that \cWB\ typically includes smaller spatial support 
in it's posterior, we come to the conclusion that \LIB\ posteriors typically 
are strongly peaked with long tails and broad spatial support. \cWB\ 
posteriors are less peaked, with more uniform probabilities and smaller 
spatial support.

Additional studies are underway to understand the statistics and 
systematics of sky maps provided by these methods, not only on injected
signals, but also on Gaussian and non-Gaussian noise artefacts. 

\subsection{Summation}

Tables \ref{table:cWB statistics} and \ref{table:LIB statistics} 
present a synopsis of the searched areas and angular offsets for 
each morphology and detector configuration. A few example posteriors 
are plotted in Section \ref{appendix:sample posteriors}.

\begin{table*}
  \caption{tabular summary of \cWB\ localization. Statistical error is on the order a few percent.}
  \begin{center}
  \begin{tabular}{p{2.0cm} r||c|c|c|c||c|c|c|c}
  \hline
  \hline
  \multicolumn{2}{c||}{year}                                                  & \multicolumn{4}{c||}{2015} & \multicolumn{4}{c}{2016} \\
  \multicolumn{2}{c||}{network}                                                 & \multicolumn{4}{c||}{HL} & \multicolumn{4}{c}{HLV} \\
  \cline{3-10}
  \multicolumn{2}{c||}{morphology}                                                     & \BBH & \SG & \G & \WNB & \BBH\ & \SG & \G & \WNB  \\
  \hline
  \hline
  \multirow{6}{2cm}{fraction (in \%) with searched area less than}  & \rl5 deg$^2$        & $3.1$ & $3.4$ & $3.1$ & $6.2$ & $6.7$ & $9.3$ & $12.0$ & $17.2$ \\
  \multirow{6}{2cm}{}  & \rl20 deg$^2$                                                    & $11.5$ & $12.7$ & $10.9$ & $17.8$ & $18.9$ & $22.2$ & $28.8$ & $32.5$ \\
  \multirow{6}{2cm}{}  & \rl100 deg$^2$                                                   & $35.3$ & $37.2$ & $37.1$ & $51.8$ & $47.3$ & $52.3$ & $54.9$ & $61.3$ \\
  \multirow{6}{2cm}{}  & \rl200 deg$^2$                                                   & $51.6$ & $52.2$ & $49.2$ & $69.7$ & $62.3$ & $66.9$ & $69.1$ & $75.8$ \\
  \multirow{6}{2cm}{}  & \rl500 deg$^2$                                                   & $75.9$ & $69.2$ & $73.8$ & $86.5$ & $82.2$ & $85.8$ & $84.9$ & $91.2$ \\
  \multirow{6}{2cm}{}  & \rl1000 deg$^2$                                                  & $89.2$ & $82.2$ & $87.1$ & $95.6$ & $93.1$ & $94.8$ & $95.0$ & $98.0$ \\
  \hline
  \multirow{6}{2cm}{fraction (in \%) with $\delta\theta$ less than}  & \rl1$^\circ$           & $1.3$ & $1.4$ & $0.8$ & $2.7$ & $3.6$ & $2.8$ & $9.8$ & $10.0$ \\
  \multirow{6}{2cm}{}  & \rl5$^\circ$                                                         & $12.9$ & $8.5$ & $11.7$ & $13.0$ & $22.6$ & $13.0$ & $29.3$ & $19.0$ \\
  \multirow{6}{2cm}{}  & \rl15$^\circ$                                                        & $37.2$ & $27.1$ & $33.2$ & $34.1$ & $37.6$ & $26.4$ & $45.8$ & $32.6$ \\
  \multirow{6}{2cm}{}  & \rl45$^\circ$                                                        & $73.1$ & $61.4$ & $66.0$ & $70.9$ & $61.3$ & $57.3$ & $67.9$ & $59.2$ \\
  \multirow{6}{2cm}{}  & \rl60$^\circ$                                                        & $79.5$ & $68.4$ & $71.1$ & $74.9$ & $66.7$ & $62.2$ & $71.7$ & $64.7$ \\
  \multirow{6}{2cm}{}  & \rl90$^\circ$                                                        & $83.1$ & $74.0$ & $75.8$ & $77.9$ & $71.4$ & $67.1$ & $74.3$ & $70.0$ \\
  \hline
  \multicolumn{2}{r||}{median searched area}                                              & 184.6 deg$^2$ & 181.6 deg$^2$ & 209.9 deg$^2$ & 93.0 deg$^2$ & 112.5 deg$^2$ & 91.7 deg$^2$ & 71.3 deg$^2$ & 61.3 deg$^2$ \\
  \multicolumn{2}{r||}{median $\delta\theta$}                                             & 23.1$^\circ$ & 31.6$^\circ$ & 25.7$^\circ$ & 23.9$^\circ$ & 27.5$^\circ$ & 36.7$^\circ$ & 18.6$^\circ$ & 33.9$^\circ$ \\
  \hline
  \end{tabular}
  \end{center}
  \label{table:cWB statistics}
\end{table*}

\begin{table*}
  \caption{tabular summary of \LIB\ localization. Statistical error is on the order of a few percent.}
  \begin{center}
  \begin{tabular}{p{2.0cm} r||c|c|c|c||c|c|c|c}
  \hline
  \hline
  \multicolumn{2}{c||}{year}                                                  & \multicolumn{4}{c||}{2015} & \multicolumn{4}{c}{2016} \\
  \multicolumn{2}{c||}{network}                                                 & \multicolumn{4}{c||}{HL} & \multicolumn{4}{c}{HLV} \\
  \cline{3-10}
  \multicolumn{2}{c||}{morphology}                                                     & \BBH & \SG & \G & \WNB & \BBH & \SG & \G & \WNB  \\
  \hline
  \hline
  \multirow{6}{2cm}{fraction (in \%) with searched area less than}  & \rl5 deg$^2$ & $1.8$ & $4.0$ & $2.3$ & $4.7$             & $8.6$ & $25.5$ & $18.5$ & $10.4$ \\
  \multirow{6}{2cm}{}  & \rl20 deg$^2$                                             & $9.4$ & $14.3$ & $7.0$ & $14.2$      & $23.1$ & $47.4$ & $43.0$ & $23.9$ \\
  \multirow{6}{2cm}{}  & \rl100 deg$^2$                                            & $31.8$ & $39.0$ & $34.8$ & $35.2$ & $53.8$ & $75.8$ & $73.6$ & $46.8$ \\
  \multirow{6}{2cm}{}  & \rl200 deg$^2$                                            & $46.8$ & $52.9$ & $49.2$ & $51.2$ & $65.7$ & $84.6$ & $84.4$ & $59.2$ \\
  \multirow{6}{2cm}{}  & \rl500 deg$^2$                                            & $70.2$ & $71.7$ & $72.7$ & $65.1$ & $82.9$ & $92.6$ & $93.3$ & $69.8$ \\
  \multirow{6}{2cm}{}  & \rl1000 deg$^2$                                           & $88.2$ & $82.6$ & $89.1$ & $74.9$ & $93.0$ & $94.9$ & $97.4$ & $76.0$ \\
  \hline
  \multirow{6}{2cm}{fraction (in \%) with $\delta\theta$ less than}  & \rl1$^\circ$  & $1.0$ & $2.1$ & $1.2$ & $2.1$ & $6.2$ & $11.4$ & $12.0$ & $5.5$ \\
  \multirow{6}{2cm}{}  & \rl5$^\circ$                                                & $8.6$ & $8.8$ & $11.7$ & $9.5$ & $34.5$ & $31.9$ & $51.4$ & $17.2$ \\
  \multirow{6}{2cm}{}  & \rl15$^\circ$                                               & $32.2$ & $25.8$ & $30.1$ & $28.6$ & $54.6$ & $53.8$ & $66.8$ & $31.3$ \\
  \multirow{6}{2cm}{}  & \rl45$^\circ$                                               & $66.8$ & $63.7$ & $63.7$ & $61.9$ & $77.1$ & $78.3$ & $83.7$ & $63.6$ \\
  \multirow{6}{2cm}{}  & \rl60$^\circ$                                               & $72.8$ & $71.0$ & $68.8$ & $67.2$ & $81.3$ & $81.8$ & $85.8$ & $70.1$ \\
  \multirow{6}{2cm}{}  & \rl90$^\circ$                                               & $77.4$ & $75.9$ & $74.2$ & $70.4$ & $83.7$ & $84.6$ & $86.5$ & $76.2$ \\
  \hline
  \multicolumn{2}{r||}{median searched area}                             & 238.5 deg$^2$ & 171.0 deg$^2$ & 208.4 deg$^2$ & 180.9 deg$^2$ & 82.5 deg$^2$ & 22.2 deg$^2$ & 31.3 deg$^2$ & 121.3 deg$^2$ \\
  \multicolumn{2}{r||}{median $\delta\theta$}                           & 26.6$^\circ$ & 29.4$^\circ$ & 27.1$^\circ$ & 30.4$^\circ$ & 11.1$^\circ$ & 13.3$^\circ$ & 4.9$^\circ$ & 27.5$^\circ$ \\
  \hline
  \end{tabular}
  \end{center}
  \label{table:LIB statistics}
\end{table*}

We should note that, because the sources of un-modeled burst
 signals may be unknown, they may not be distributed 
uniformly in volume. If one takes an agnostic view, they may
 use only the maximum likelihood estimate without our 
effective prior. For our injection set, this corresponds to 
an increase in the median searched areas because the algorithm 
no longer searches the ring according to the 
antenna patterns. We tested this using \cWB\ and see that the 
median searched areas are at least 25\% larger (2015 \G) without
the effective prior, and as much as 59\% larger (2015 
\SG). 

Furthermore, we expect \BBH\ signals to be circularly polarized. 
Incorporating this information into the search should improve 
the localization. We tested this with \cWB\ and observed no 
change in the 2015 data set. This is because the two-detector 
regulator values force the algorithm to reconstruct signals 
with a single polarization, and knowledge that the signal 
is circularly polarized adds nothing. 
However, in the three-detector network, we see that the median 
searched area improves by nearly a factor of 3, from 112.5 deg$^2$ 
to 38.0 deg$^2$. With the polarization constraint, this is comparable
to the localization observed with lighter systems using matched
filtering techniques [~\cite{Singer:2014}]. However, ~\cite{Singer:2014}
focus on different sources than our \BBH\ injection set and a
direct comparison between our results requires careful consideration.

Customizing gravitational-wave searches to specific source models 
may improve our ability to localize them in the sky in a significant way.

\section{Systematics}\label{section:systematics}

We have investigated several sources of potential systematics 
in \cWB's and \LIB's reconstructed sky positions. 
These can be roughly classified
as ``calibration'' and ``accuracy'' biases. 

\subsection{Posterior calibration}\label{section:pp}

We first study systematics associated with the probability 
assigned by the localization algorithms on a pixel-by-pixel 
basis.
This can be considered a ``calibration'' problem with 
the posterior distributions over the sky. On average, we 
expect the confidence region containing N\% cumulative 
probability to contain N\% of the detected 
injections. 

For \cWB\ we find the calibration of the posterior distributions
to depend strongly on the choice of regulators, in addition to
its dependance on the intrinsic parameters of the simulated
events. The dependance on the regulators seems to dominate these
systematics. We have seen variation in
frequentist confidence levels
as large as 50\% in both directions,
i.e., undercovering and overcovering.
For the specific
choice of regulators (Section \ref{section:algorithmic params}), 
\cWB\ generally  underestimates the actual confidence 
at low confidence levels and overestimates the actual confidence at 
very high confidence levels. \cWB's 50\% confidence regions contain 
between 65\%-85\% of the detected signals in 2015 and between 
65\%-75\% of detected signals in 2016, depending on morphology. 
Because the choice of regulators will depend on the
character of real data,
systematics and the regulators'
impact on localization and detection efficacy need to be
re-evaluated when advanced LIGO and Virgo will come online.
This may include recalibration to account for observed systematics.

\LIB\ also shows calibration issues.
Its 50\% confidence regions contain between 45\%-55\% of 
the detected signals in 2015  and between 35\%-45\% of detected 
signals in 2016, depending on morphology. 
\LIB's systematic overestimation of
the actual confidence observed in the 2016 data is due 
to differences in the population of detected events and the
population expected by  \LIB's priors.
We have verified that when detected events sample \LIB's
uniform-in-volume prior, \LIB's posteriors are properly
calibrated.
Selection effects in events detected by \cWB\ may result in such deviations
from the assumed prior. 

Such calibration issues are not particular to burst searches,
and are observed with localization pipelines targeting binary
neutron star coalescences [~\cite{Sidery:2014}].
Although these systematics warrant further investigation,
they do not imply that these algorithms cannot be used
to direct electromagnetic follow-up of gravitational wave events.
Any observing plan consists of a set of fields sorted by an 
ordinal function of their probability to contain signal. 
The observing strategy will not depend on the actual function, 
as long as it preserves ordering.

\subsection{Bias in reconstructed positions}
\label{section:est bias}

We also compared the distribution of injected positions for detected events 
against their estimated positions. We expect these distributions to match 
over an ensemble average, implying the algorithms typically localize
signals to the same regions of the sky from which they are detected.
The injected positions of detected events always follow the antenna patterns of our 
network of detectors. We therefore expect the reconstructed positions to similarly
follow the antenna patterns.

Both algorithms behave essentially as expected in the two-detector case
(2015). 
However, we observe a bias in \cWB's estimated positions 
in the
three-detector network (2016).
As with the calibration of the posteriors
(Section ~\ref{section:pp} ), the size and direction of this bias
depends strongly on the choice of regulators.
The estimated sky positions may be clumped into regions with different shapes
than the injected distributions and may not coincide with the maxima of the
antenna patterns.
For the regulators used in this study,
this offset corresponds to between
20$^\circ$ and 40$^\circ$.
This bias is also imprinted on the entire posterior and not just the 
maxima.
We also note that, when applying a circular polarization constraint
in \cWB, the median searched area improves and the bias is removed.
We do not observe any such bias 
with \LIB, which may partially account for the increased disagreement 
with \cWB\ in the 2016 data.

\section{Conclusions}
\label{section:conclusion}

We present a study of gravitational-wave source localization
capabilities for un-modeled signals during the early advanced 
LIGO and Virgo detector era.
In particular, we focus on the transition from 
two operational detectors in 2015 to three operational detectors 
in 2016, and quantify the improvement in localization 
associated therewith.
In performing this study we used two different localization 
algorithms: Coherent WaveBurst (\cWB), a low-latency maximum 
likelihood algorithm, and LALInferenceBurst (\LIB), a Markov 
chain Monte Carlo (MCMC) parameter estimation algorithm.
We used four signal morphologies to explore a wide range of 
possible signals detectable by generic burst searches.

We find that, while there is some variation with waveform 
morphologies, 
50\% of injected signals would be imaged after observing 100-200 deg$^2$ with two detectors
in 2015. We find that \cWB\ can reduce this to within 60-110 deg$^2$ 
in 2016 with low latency, and \LIB\ may reduce the median searched area 
to as little as 22 deg$^2$ when the signal matches the template.
Tables \ref{table:cWB statistics} and \ref{table:LIB statistics} 
summarize our findings. While the searched areas may be small,
we should remember that the posterior may be spread across large
portions of the sky, including antipodal points.

Importantly, we also introduce an effective prior on the source 
position due to anisotropic antenna patterns and knowledge of 
signal amplitude distributions.
With this prior, we find 
\cWB\ performs comparably to the full MCMC \LIB\ algorithm 
in the two-detector configuration (2015) with significantly
lower latency.
This is true even for signal morphologies
that correspond to \LIB's templates.
However, \LIB\ significantly 
improves upon rapid localizations provided by \cWB\ with
three detectors (2016) for most of the considered 
morphologies.

Furthermore, we find that \cWB\ systematically localizes \WNB\ signals more 
accurately than \LIB. This is because \cWB\ makes no assumption on the signal
morphology while \LIB\ assumes a single \SG\ template.
\cWB\ localizes \WNB\ signals better than \SG\ or \G\ signals 
because of differences in the signal morphologies. We expect 
high-frequency or large bandwidth signals to be the better 
localized than low-frequency or narrow bandwidth signals.

We also studied the posteriors produced by \cWB\ and \LIB\ on 
and event-by-event basis. We found that the two agorithms agree 
on the maximum a posteriori points to a remarkable degree 
in 2015, typically to within $8^\circ$. 
We find that the selected pixels agree to over 50\% in the 
two-detector network, but agree significantly less in the 
three-detector network.
Several considerations lead us to the heuristic conclusion 
that \LIB\ posteriors are typically sharply peaked with long 
tails and large spatial support. \cWB\ posteriors are typically 
less peaked, more uniform with smaller spatial support.

Finally, we investigate and quantify some systematics with 
both methods. Both \cWB\ and \LIB\ show calibration issues 
with their posteriors, in that the Bayesian confidence regions
 do not correspond to their frequentist counterparts.
\cWB's regulators introduce biases in the reconstructed 
positions in the 2016 data set, and in a small fraction of 
events modulate the posterior so strongly that the source's 
location is outside of the posterior's support.

We expect templated searches targeting known signal morphologies 
to localize signals more accurately than generic un-modeled 
searches. When compared to an analogous study focusing on 
Binary Neutron Star (BNS) coalescences [~\cite{Singer:2014}], we see that the 
localization of BNS signals is indeed more accurate than generic 
bursts. However, the generic burst 
searches produce median searched areas that are only 
a factor of 2-3 larger than BNS median searched areas. In fact, the median searched areas may be 
comparable if the actual burst waveform is known reasonably well in the
three-detector network, such as \LIB\ recovering \SG\ injections. For circularly polarized signals, 
\cWB\ can achieve comparable results
if it assumes the signal is circularly polarized.
While this study and ~\cite{Singer:2014} both use populations
distributed uniformly in volume, we investigate very different 
signal morphologies. Furthermore, ~\cite{Singer:2014} estimates 
the detectors duty cycles in 2016 and occasionally detect
events with only two detectors instead of all three. This could
cause our estimates to seem more similar than they really are. However, we also
use a lower false-alarm rate threshold than ~\cite{Singer:2014}, which
will increase our error estimates systematically because we will include quieter events.
Any direct comparison between these studies should include careful 
consideration of such differences.

\Rev{
Importantly, we find that telescope networks attempting to follow up BNS 
events from advanced gravitational wave detectors by searching large error areas
can search the comparably sized areas for burst events.
}
Given 
the nature of such generic events, electromagnetic observations 
will be instrumental in placing gravitational wave observations 
in an astrophysical context.

\section*{Acknowledgements}

The authors would like to thank J. Veitch and A. Vecchio for 
comments and suggestions about \LIB, W. Farr for his script 
to convert MCMC posterior samples into a pixelated posterior
as well as R. Vaulin and R. Lynch 
for many useful discussions throughout the course of this 
research. The authors also acknowledge L. Price for helpful comments 
when preparing this manuscript and L. Singer for creating 
the 2015 Gaussian noise data.
LIGO was constructed by the California Institute of Technology and 
Massachusetts Institute of Technology with funding from the NSF and 
operates under cooperative agreement PHY-0757058. This work was also 
supported from NSF awards PHY-1205512 and PHY-0855313 to the University of Florida.

\bibliography{refs}

\newpage
\section{Appendix}

\subsection{Injection parameter ranges}

Table \ref{table:params} lists the injection parameter ranges 
and distributions used. Definintion of the waveform morphologies 
are provided in Section \ref{section:Data preparation and injection generation}.

\begin{table*}[ht]
  \caption{Injection population parameters}
  \label{table:params}
  \begin{center}
  \begin{tabular}{c|c|c|c|c|c}
    \hline
    \hline
    \multicolumn{3}{c|}{} & minimum & maximum & distribution \\
    \hline
    \hline
    \multirow{6}{*}{\SG} & \multicolumn{2}{c|}{$f_o$}   & $40\, Hz$  & $1500\, Hz$ & \multirow{4}{*}{$\mathrm{d} n \propto constant$} \\
    \multirow{6}{*}{}    & \multicolumn{2}{c|}{$Q=\sqrt{2}\pi\tau f_o$} & 3 & 30 & \multirow{4}{*}{} \\
    \multirow{6}{*}{}    & \multicolumn{2}{c|}{$\cos\left(\alpha\right)$} & $0$ & $1$ & \multirow{4}{*}{} \\
    \multirow{6}{*}{}    & \multicolumn{2}{c|}{$\phi_o$} & $0$ & $2\pi$ & \multirow{4}{*}{} \\
    \cline{2-6}
    \multirow{6}{*}{}    & \multirow{2}{*}{$h_{rss}$} &  2015 & $3.0000\cdot10^{-23}\, Hz^{-1/2}$ & $1\cdot10^{-15}\, Hz^{-1/2}$ & \multirow{2}{*}{$\mathrm{d}n \propto D_L^2\mathrm{d}D_L \propto h_{rss}^{-4}\mathrm{d}h_{rss}$} \\
    \multirow{6}{*}{}    & \multirow{2}{*}{}          &  2016 & $2.0625\cdot10^{-23}, Hz^{-1/2}$ & $1\cdot10^{-15}\, Hz^{-1/2}$ & \multirow{2}{*}{} \\
    \hline
    \hline
    \multirow{4}{*}{\G}  & \multicolumn{2}{c|}{$\tau$} & $1\,ms$ & $10\,ms$ & \multirow{3}{*}{$\mathrm{d} n \propto constant$} \\
    \multirow{4}{*}{}    & \multicolumn{2}{c|}{$\cos\left(\alpha\right)$} & $0$ & $1$ & \multirow{3}{*}{}\\
    \cline{2-6}
    \multirow{4}{*}{}    & \multirow{2}{*}{$h_{rss}$} & 2015 & $4.0000\cdot10^{-23}\, Hz^{-1/2}$ & $1\cdot10^{-15}\, Hz^{-1/2}$ & \multirow{2}{*}{$\mathrm{d}n \propto h_{rss}^{-4} \mathrm{d} h_{rss}$} \\
    \multirow{4}{*}{}    & \multirow{2}{*}{ }         & 2016 & $2.7500\cdot10^{-23}\, Hz^{-1/2}$ & $1\cdot10^{-15}\, Hz^{-1/2}$ & \multirow{2}{*}{} \\
    \hline
    \hline
    \multirow{5}{*}{\WNB}    & \multicolumn{2}{c|}{$\tau$} & $5\,ms$ & $100\,ms$ & \multirow{3}{*}{$\mathrm{d} n \propto constant$} \\
    \multirow{5}{*}{}    & \multicolumn{2}{c|}{$f_o \sim (f_{max} + f_{min})/2$} & $40\,Hz$ & $1500\,Hz$ & \multirow{3}{*}{} \\
    \multirow{5}{*}{}    & \multicolumn{2}{c|}{$\sigma_f \sim (f_{max}-f_{min})/2$} & $10\,Hz$ & $500\,Hz$ & \multirow{3}{*}{} \\
    \cline{2-6}
    \multirow{5}{*}{}    & \multirow{2}{*}{$h_{rss}$} & 2015 & $4.0000\cdot10^{-23}\, Hz^{-1/2}$ & $1\cdot10^{-15}\, Hz^{-1/2}$ & \multirow{2}{*}{$\mathrm{d}n \propto h_{rss}^{-4} \mathrm{d} h_{rss}$} \\
    \multirow{5}{*}{}    & \multirow{2}{*}{}          & 2016 &  $2.7500\cdot10^{-23}\, Hz^{-1/2}$ & $1\cdot10^{-15}\, Hz^{-1/2}$ & \multirow{2}{*}{} \\
    \hline
    \hline
    \multirow{6}{*}{\BBH} & \multicolumn{2}{c|}{$M_1$} & 15 $M_\odot$ & 25 $M_\odot$ & \multirow{4}{*}{$\mathrm{d}n \propto constant$} \\
    \multirow{6}{*}{}     & \multicolumn{2}{c|}{$M_2$} & 15 $M_\odot$ & 25 $M_\odot$ & \multirow{4}{*}{} \\
    \multirow{6}{*}{}     & \multicolumn{2}{c|}{$S_1$} & 0.0          & 0.9          & \multirow{4}{*}{} \\
    \multirow{6}{*}{}     & \multicolumn{2}{c|}{$S_2$} & 0.0          & 0.9          & \multirow{4}{*}{} \\
    \cline{2-6}
    \multirow{6}{*}{}     & \multirow{2}{*}{$z$} & 2015 & $10^{-4}$ & $0.2218$ & \multirow{2}{*}{$\mathrm{d}n \propto \frac{\mathrm{d}V_{comov}}{\mathrm{d}z}\mathrm{d}z$} \\
    \multirow{6}{*}{}     & \multirow{2}{*}{}    & 2016 & $10^{-4}$ & $0.33$   & \multirow{2}{*}{} \\
    \hline
    \hline
  \end{tabular}
  \end{center}
\end{table*}

\subsection{Algorithmic parameters}
\label{section:algorithmic params}

We used version 3481M of the cWB repository,
searching for signals with 
frequencies betwen 32-2048 Hz. Our detection thresholds 
were set to $\rho=8.0$, netcc$=0.7$. While false-alarm 
rates will depend on the noise in our detectors, these 
detections correspond to a false-alarm-rate of 1 yr$^{-1}$ in historical 
non-Gaussian noise. Table \ref{table:cWB params} lists the 
regulator values used in this study, which we varied 
depending on the detector network. We note that the regulators 
in 2015 force \cWB\ to reconstruct a single polarization, 
while the regulators in 2016 are almost, but not quite, 
turned off. The consequences of $\delta\neq0$ in 2016 
are discussed in Section \ref{section:systematics}

\begin{table*}[ht]
  \caption{CoherentWaveBurst search parameters.}
  \label{table:cWB params}
  \begin{center}
  \begin{tabular}{c||c|c|c|c|c|c|c|c}
    \hline
    \hline
    year/network & SVN revision No. & cWB version & $f_{low}$ & $f_{high}$ & $\rho$ & netCC & $\delta$ & $\gamma$ \\
    \hline
    \hline
    2015/HL  & \multirow{2}{*}{3481M} & \multirow{2}{*}{2G} & \multirow{2}{*}{32 Hz} & \multirow{2}{*}{2048 Hz} & \multirow{2}{*}{8.00} & \multirow{2}{*}{0.70} & $-10^{40}$ & $0.5$ \\
    2016/HLV & \multirow{2}{*}{}     & \multirow{2}{*}{}   & \multirow{2}{*}{}      & \multirow{2}{*}{}        & \multirow{2}{*}{}     & \multirow{2}{*}{}     & $0.05$     & $0.5$ \\
    \hline
  \end{tabular}
  \end{center}
\end{table*}

We ran \LIB\ with three parallel chains with 500 live points each, and estimated noise power spectral densities from the data separately for each event using 96 seconds near the trigger time. Table \ref{table:LIB params} list the prior ranges used for \LIB's \SG\ template.

\begin{table*}[ht]
  \caption{LALInferenceBurst prior ranges}
  \label{table:LIB params}
  \begin{center}
  \begin{tabular}{c|c|c|c|c|c}
    \hline
    \hline
        & $\log\left(h_{rss}\right)$ & $f_o$ & $Q$ & $\alpha$ & $\phi$ \\
    \hline
    \hline
    minimum & -53.0 & 1 Hz & 2 & 0 & 0 \\
    maximum & -46.5 & 1300 Hz & 35 & 2$\pi$ & 2$\pi$ \\
    distribution & $p(h_{rss}) \propto h_{rss}^{-4}$ & $p(f_o)\propto$ constant & $p(Q)\propto$ constant & $p(\alpha)\propto$ constant & $p(\phi)\propto$ constant \\
    \hline
  \end{tabular}
  \end{center}
\end{table*} 


\subsection{Derivation of effective priors}
\label{appendix:effective priors}

We can write down an astrophysically motivated prior, such as 
a uniform in co-moving volume distribution. However, for burst 
signals, we do not immediately have a good estimate for the 
distance ($D$) or the energy scale ($E$). We can relate this 
to the observed data ($h$) through

\begin{equation}
\frac{E}{D^2} \propto \int\mathrm{d}f\, f^2 \left(\left|h_+\right|^2 + \left|h_\times\right|^2\right)
\end{equation}

To obtain a prior on $h$, we should marginalize over all 
possible $D$ and $E$.

\begin{eqnarray}
p(h,E,D)\mathrm{d}h\,\mathrm{dE}\,\mathrm{dD} & = & p(h|E,D)\mathrm{d}h \cdot p(E)\mathrm{d}E \cdot p(D)\mathrm{d}D \\
                                          & \propto & \delta\left(h-\lambda\sqrt{\frac{E}{D^2}}\right)\mathrm{d}h \cdot p(E)\mathrm{d}E \cdot D^2 \mathrm{d}D \\
\end{eqnarray}

where $h^2=\int\mathrm{d}f\, f^2 \left(\left|h_+\right|^2 + \left|h_\times\right|^2\right)$ 
and $\lambda$ is a proportionality constant. Marginalization yields

\begin{eqnarray}
p(h) & \propto & \int\mathrm{d}E\,p(E)\int\mathrm{d}D\, D^2 \delta\left(h-\lambda\sqrt{\frac{E}{D^2}}\right) \\
     & \propto & \int\mathrm{d}E\,p(E)\int\mathrm{d}D\, D^2 \delta\left(D-\lambda\sqrt{\frac{E}{h^2}}\right)\frac{\lambda\sqrt{E}}{h^2} \\
     & \propto & h^{-4} \lambda^3 \int\mathrm{d}E\,p(E) E^{3/2} \\
     & \propto & h^{-4}
\end{eqnarray}

Because we do not know the actual waveform a priori, we must 
marginalize over the waveform to compute the posterior. With
 a small number of detectors, the likelihood is not strongly
 peaked around the maximum likelihood estimate and we can 
approximate 

\begin{equation}
\int\mathrm{d}h\, p(h) \sim h_{ML}^{-3}
\end{equation}

where $h_{ML}$ is the maximum likelihood reconstructed signal. 
This can be approximated by 
$h_{ML} \sim \left(F_+^2 + F_\times^2\right)^{-1/2}$, 
where $F_{+,\times}$ are the antenna patterns for the entire 
network in the dominant polarization frame 
[~\cite{Klimenko:2011,Sutton:2010}]. We then expect the effective 
prior on ($\theta,\phi$) to be something like

\begin{equation}
p_{\mathrm{eff}}(\theta,\phi) = \left(F_+^2 + F_\times^2\right)^{3/2}
\end{equation}

We observe that this prior improves source localization, although 
there is only a weak dependence on the actual exponent used. For 
most signals, any positive power of the antenna patterns tends 
to order the pixels in the correct way to reduce the searched 
area. 
However, the posterior's calibration depends strongly on this exponent. 
\cWB's regulators skew it's calibration and the two affects are difficult 
to separate.

\subsection{Sample posteriors}
\label{appendix:sample posteriors}

\begin{figure*}
	\begin{center}
	\begin{minipage}{0.4\textwidth}
		\begin{center}
			\includegraphics[width=1.0\textwidth, clip=True, trim=0.25cm 1.0cm 0.25cm 2.6cm]{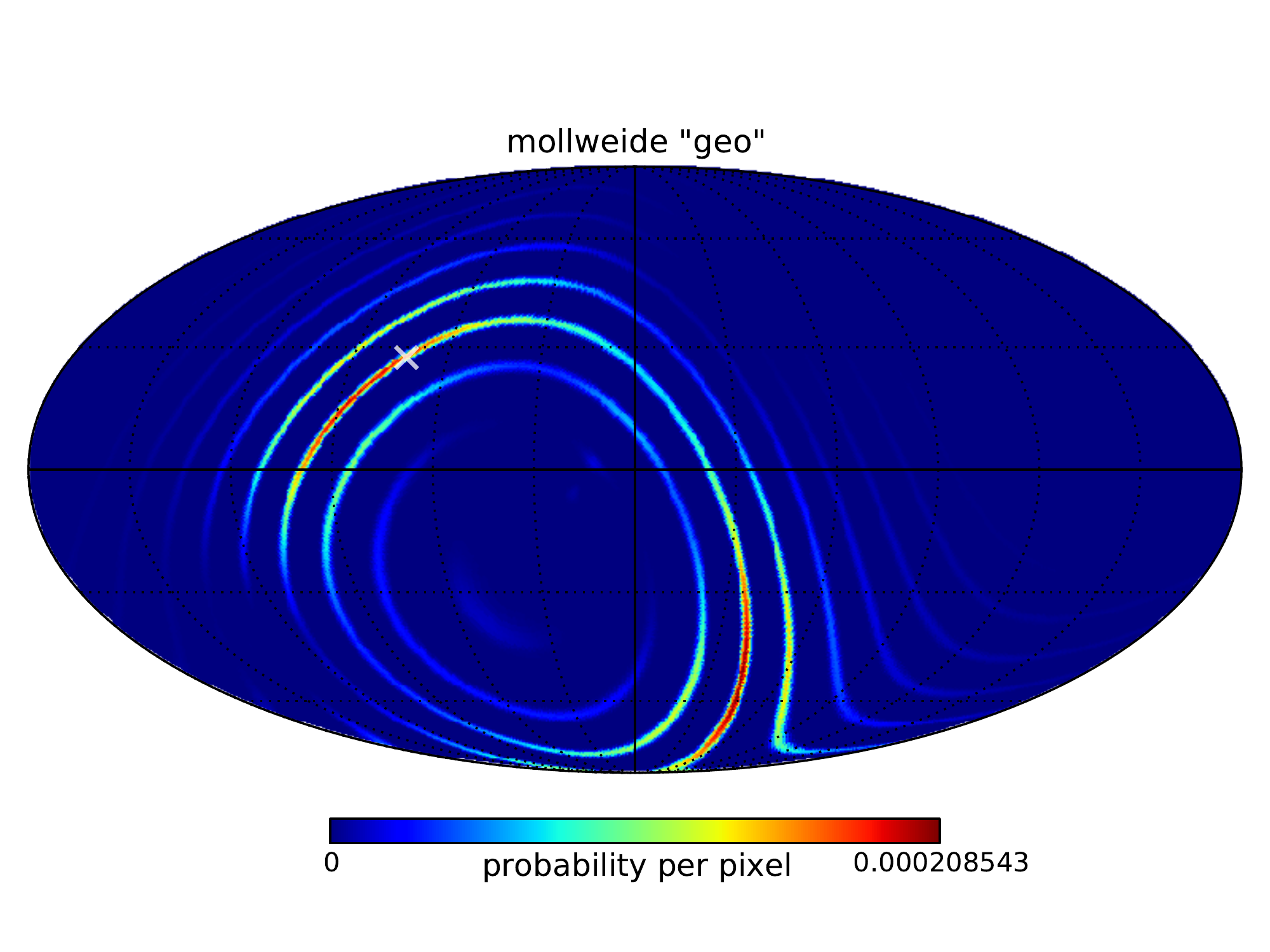} \\ 
			(a) \cWB\ 2015 \WNB\\ 
			\includegraphics[width=1.0\textwidth, clip=True, trim=0.25cm 1.0cm 0.25cm 2.6cm]{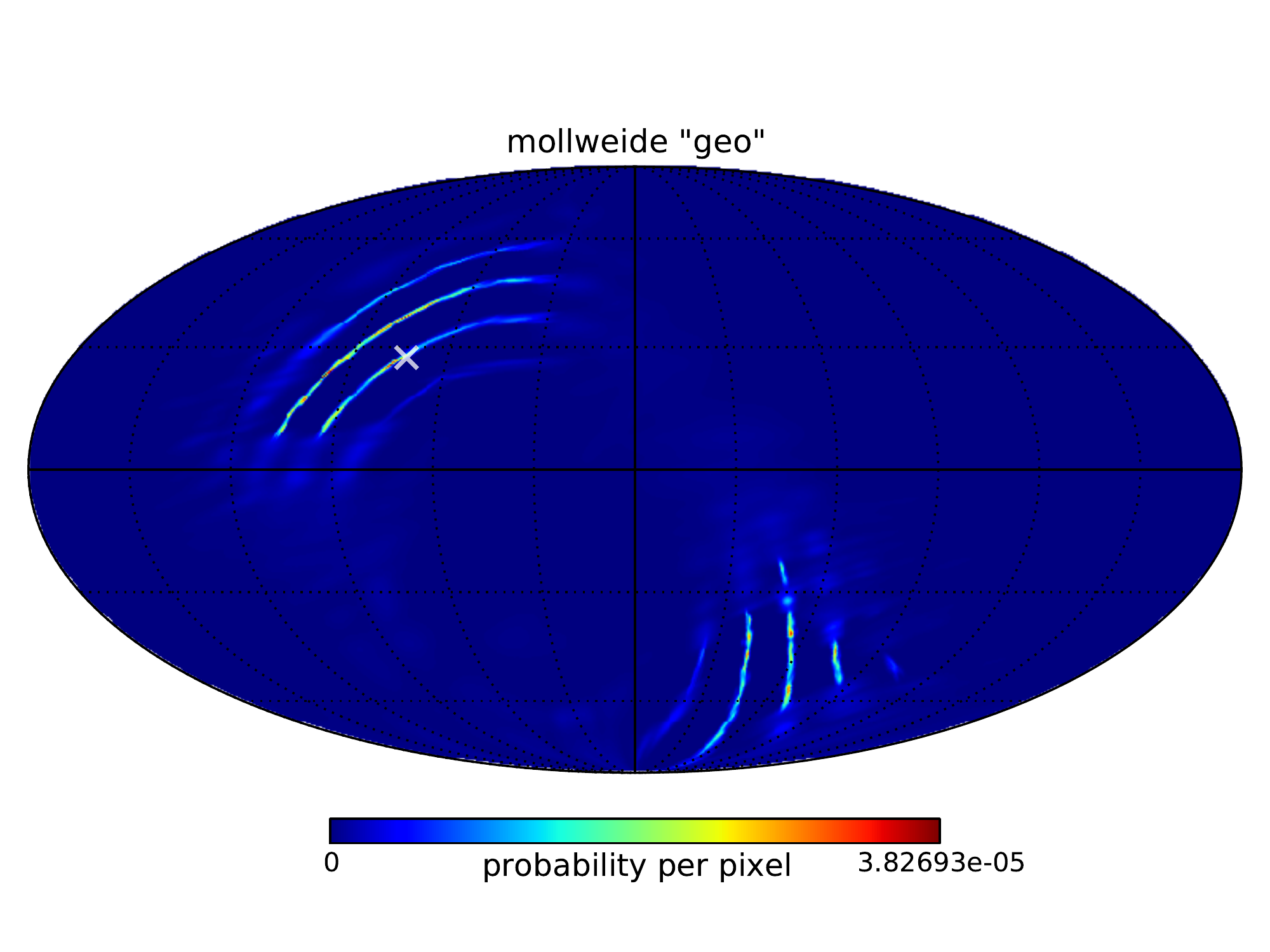} \\ 
                        (c) \LIB\ 2015 \WNB\\
		\end{center}
	\end{minipage}
	\begin{minipage}{0.4\textwidth}
		\begin{center}
                        \includegraphics[width=1.0\textwidth, clip=True, trim=0.25cm 1.0cm 0.25cm 2.6cm]{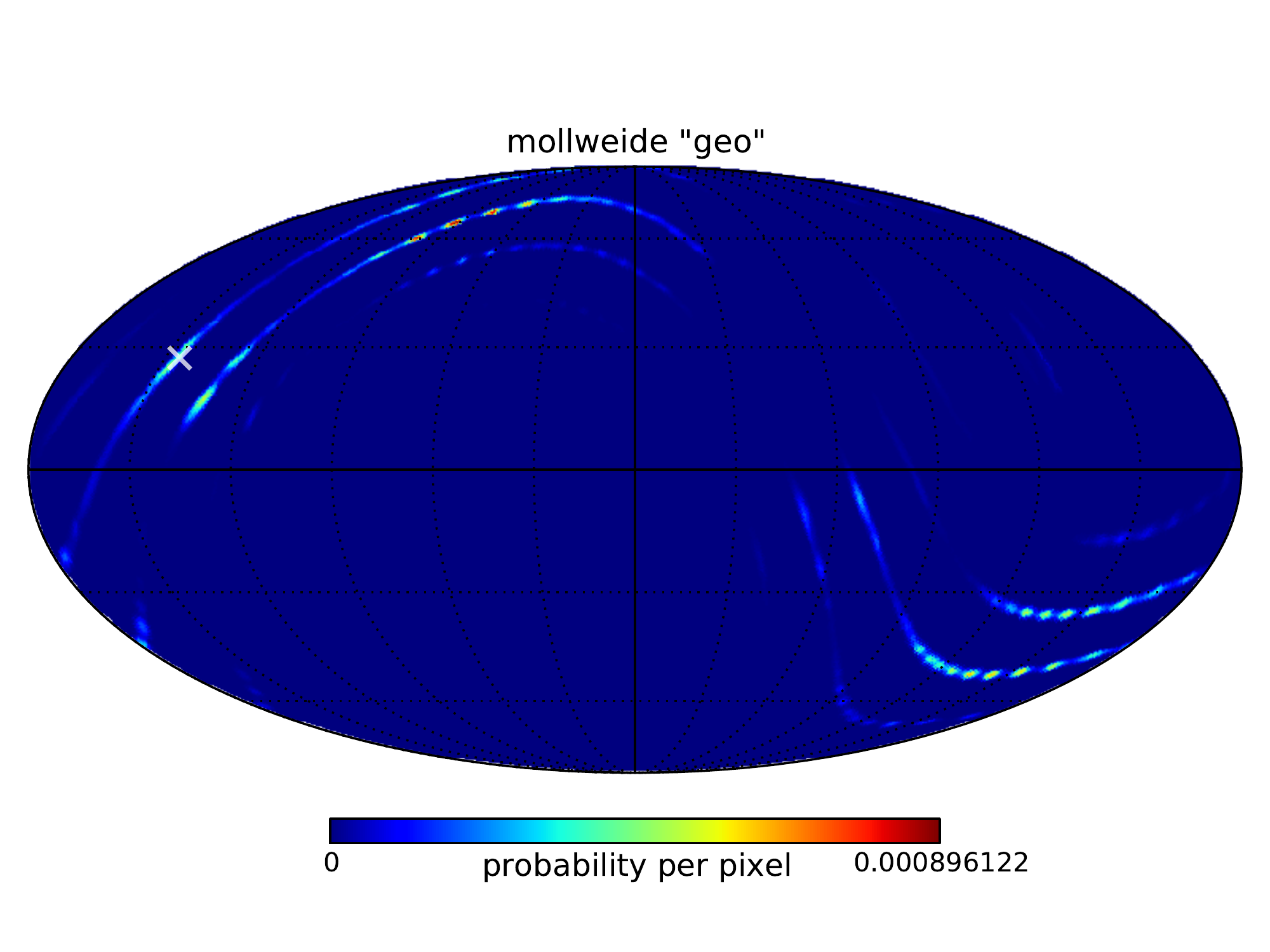} \\
                        (b) \cWB\ 2016 \WNB\\
                       \includegraphics[width=1.0\textwidth, clip=True, trim=0.25cm 1.0cm 0.25cm 2.6cm]{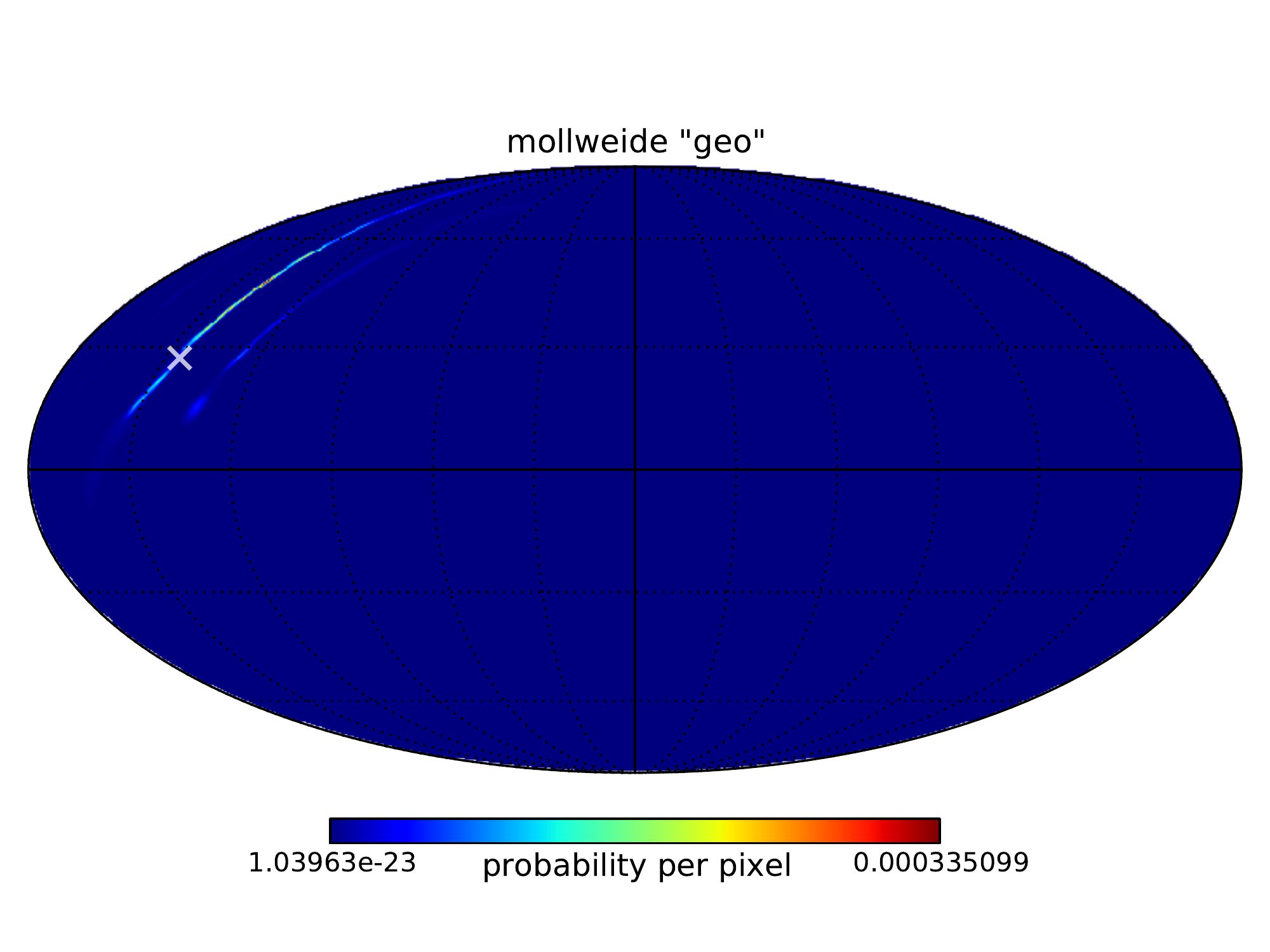} \\ 
                        (d) \LIB\ 2016 \WNB\\

		\end{center}
	\end{minipage}
	\end{center}
	\caption{Sample \WNB\ posteriors. (a,c) 2015. (b,d) 2016. (a,b) \cWB. (c,d) \LIB. An ``x'' marks the injected location. These are mollweide ``geo'' projections. The 2015 injection had SNRs of 8.81, 11.72 and 14.66 for LHO, LLO, and the entire network, respectively. The 2016 injection had SNRs of 13.02, 10.66, 1.65 and 16.91 for LHO, LLO, Virgo and the entire network, respectively.} 
        \label{figure:WNB samples}
\end{figure*}

Figure \ref{figure:WNB samples} demonstrates posterior distributions produced by both \cWB\ and \LIB\ for \WNB\ injections. In the two-detector examples, we can clearly see fringe peaks characteristic of high-frequency, low-bandwidth signals. The two-detector injection has a central frequency of $\sim560$ Hz, which corresponds to an angular scale of $\sim10^\circ$. The three-detector injection has a central frequency of $\sim350$ Hz, which corresponds to $\sim16^\circ$ between the two LIGO detectors and $\sim6^\circ$ between the LIGO's and Virgo. We also see that \LIB\ and \cWB\ agree, but \LIB's posterior is ``fuzzy,'' which we expect due to template mismatch. The three-detector network shows similar fringe structures, although now we can see the main triangulation ring from the two LIGO detectors modulated by information from Virgo. In the \LIB\ posterior, we can even see the regular lattice imprinted on neighbouring triangulation rings. We note that the three-detector posteriors are less similar than the two-detector posteriors, which we discuss in detail in Sections \ref{section:comparison} and \ref{section:systematics}.

\begin{figure*}
	\begin{center}
        \begin{minipage}{0.4\textwidth}
                \begin{center}
                        \includegraphics[width=1.0\textwidth, clip=True, trim=0.25cm 1.0cm 0.25cm 2.6cm]{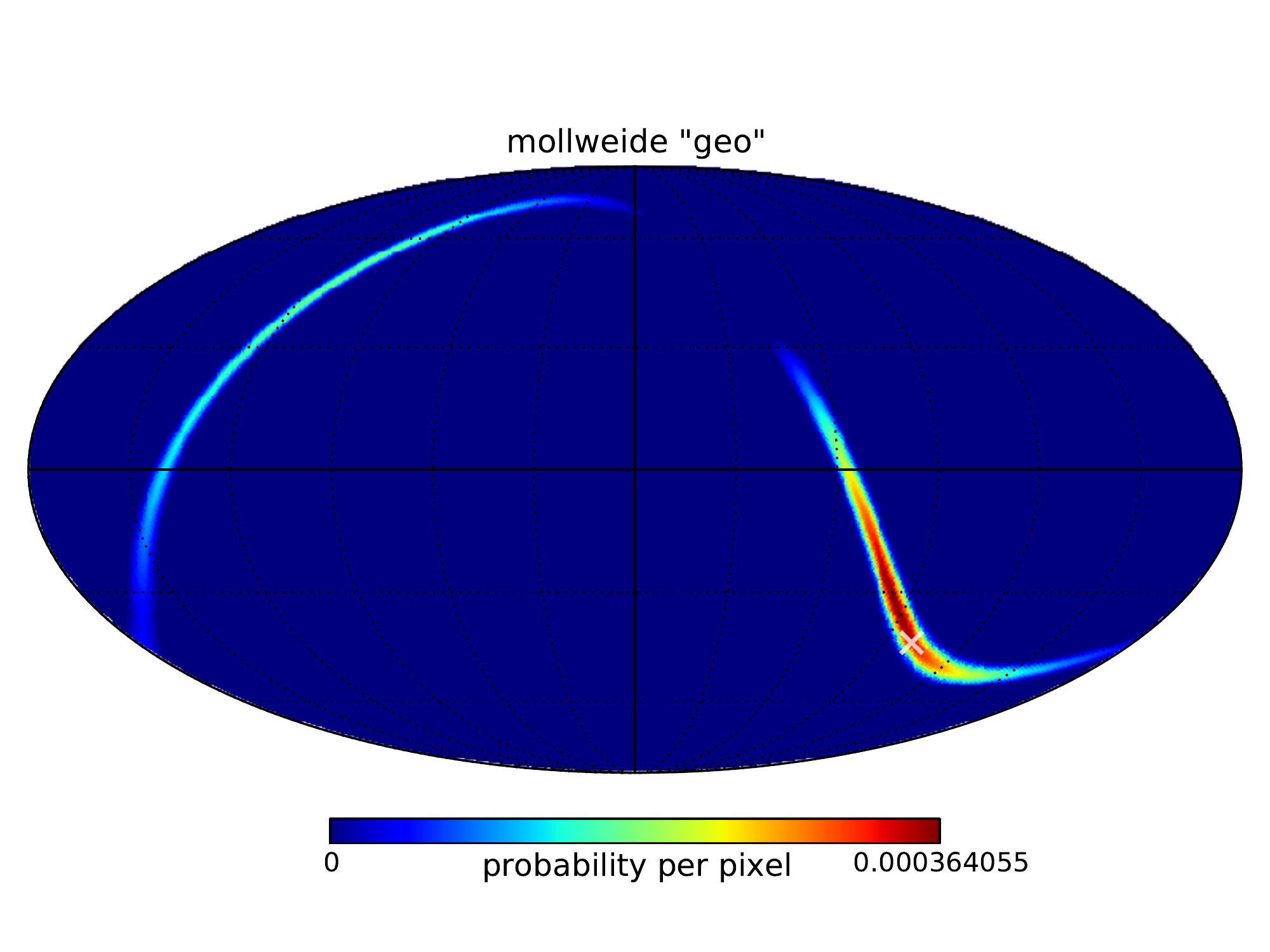} \\ 
                        (a) \cWB\ 2015 \BBH\\ 
                       \includegraphics[width=1.0\textwidth, clip=True, trim=0.25cm 1.0cm 0.25cm 2.6cm]{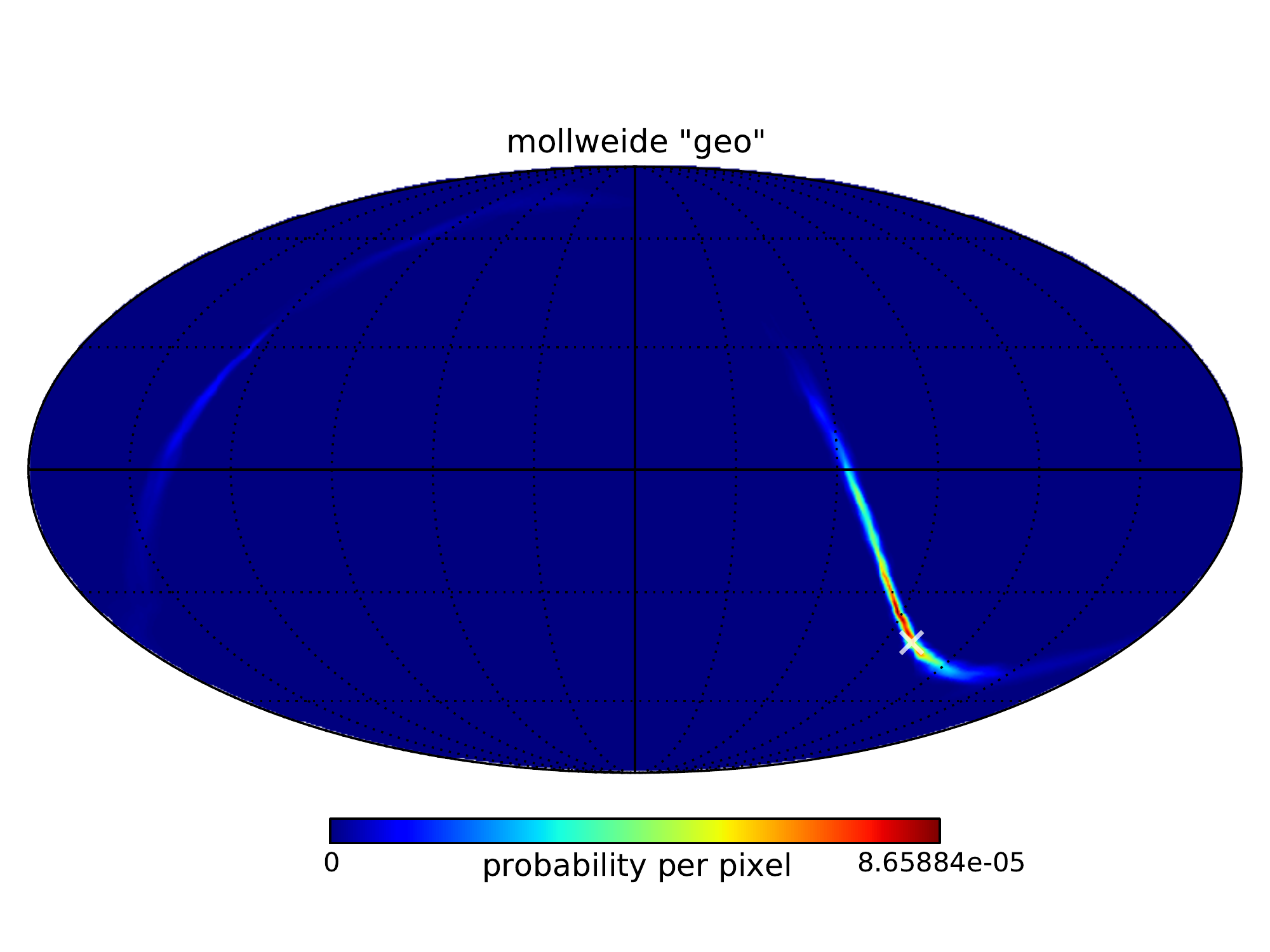} \\ 
                        (c) \LIB\ 2015 \BBH\\ 
                \end{center}
        \end{minipage}
        \begin{minipage}{0.4\textwidth}
                \begin{center}
                        \includegraphics[width=1.0\textwidth, clip=True, trim=0.25cm 1.0cm 0.25cm 2.6cm]{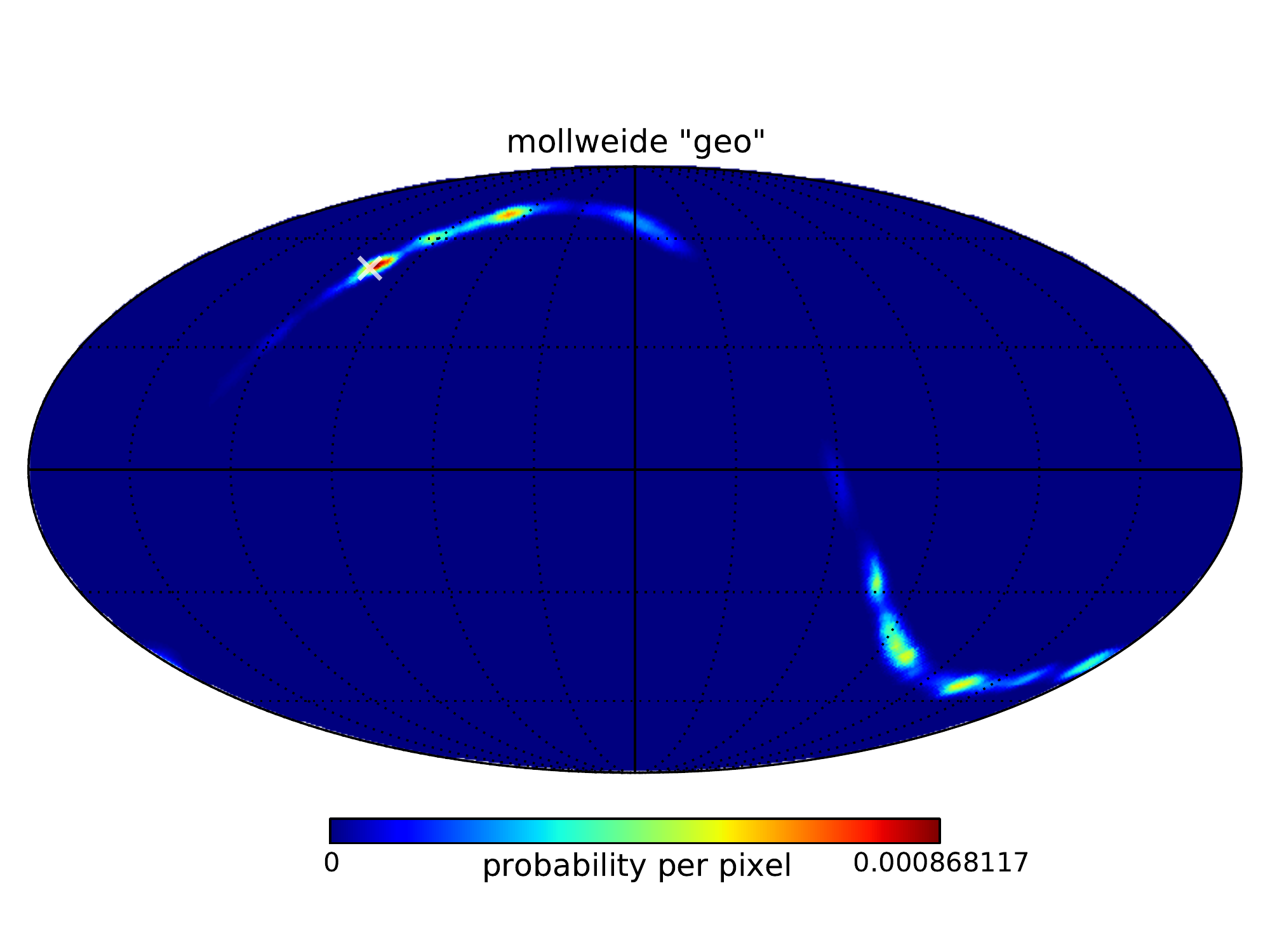} \\ 
                        (b) \cWB\ 2016 \BBH\\ 
                       \includegraphics[width=1.0\textwidth, clip=True, trim=0.25cm 1.0cm 0.25cm 2.6cm]{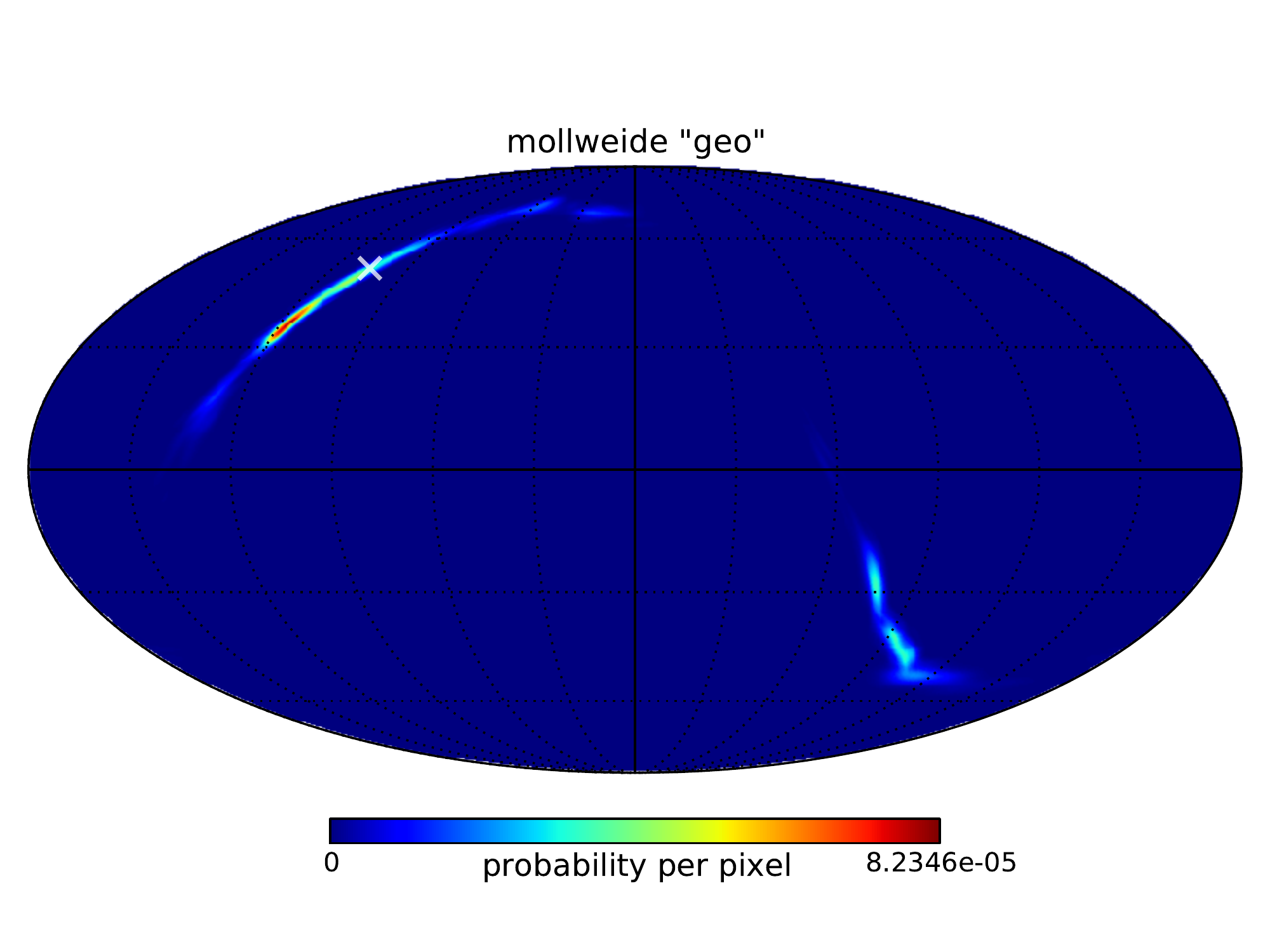} \\ 
                        (d) \LIB\ 2016 \BBH\\ 

                \end{center}
        \end{minipage}
	\end{center}
	\caption{Sample \BBH\ posteriors. (a,c) 2015. (b,d) 2016. (a,b) \cWB. (c,d) \LIB. An ``x'' marks the injected location. These are mollweide ``geo'' projections. The 2015 injection had SNRs of 11.37, 12.72, and 17.06 for LHO, LLO, and the entire network, respectively. The 2016 injection had SNRs of 14.06, 13.40, 1.85 and 19.51 for LHO, LLO, Virgo and the entire network, respectively.} 
	\label{figure:BBH samples}
\end{figure*}

Figure \ref{figure:BBH samples} is analogous to Figure \ref{figure:WNB samples}, except it shows data from \BBH\ injections. For \BBH\ injections, which have wide bandwidths relative to their central frequencies, we do not expect fringe peaks. The modulation in the three-detector network along the single triangulation ring is caused by Virgo. The \cWB\ posteriors clearly demonstrate the increase in the number of disjoint regions when progressing from the two-detector network to the three-detector network. 

\end{document}